\documentclass[twocolumn,prb,superscriptaddress,aps,10pt,bibnotes,amsfonts,amsmath,amssymb,floatfix]{revtex4-2} 

\usepackage{graphicx}
\usepackage{ifthen}
\usepackage{xcolor}
\usepackage{bm}
\usepackage{braket}
\usepackage{multirow}
\usepackage{hyperref}
\usepackage{soul}
\usepackage{svg}

\usepackage{geometry}           
\usepackage{amsmath,amssymb}
\usepackage{braket}
\graphicspath{{./}}
\usepackage{booktabs}
\usepackage{float}

\hypersetup{%backref=true,
 pdfnewwindow=true, colorlinks=true,
 linkcolor=blue, anchorcolor=blue,
 citecolor=blue, filecolor=blue,
 menucolor=blue, urlcolor=blue}

\newcommand{\beq}{\begin{equation}}
\newcommand{\eeq}{\end{equation}}
\newcommand{\red}[1]{{\color{black}#1}}

\newcommand{\fref}[1]{Fig.\,\ref{#1}}

\newcommand{\tref}[1]{Table\,\ref{#1}}

\newcommand{\eref}[1]{Eq.\,(\ref{#1})}

\newcommand{\sref}[1]{Sec.\!~\ref{#1}}

\newcommand{\appref}[1]{Appendix~\ref{#1}}

\usepackage{color}
\usepackage[normalem]{ulem}
\newlength{\figwidth}
\newcommand{\cref}[1]{Ref.\,\cite{#1}}

\begin{document}

\title{Quasiparticle GW for superconductors: \red{Toward a unified treatment of electron-phonon and electron-plasmon couplings}}
\author{Catalin D. Spataru}\email{cdspata@sandia.gov}
\affiliation{Sandia National Laboratories, Livermore, CA 94551, USA}
\author{Christopher Renskers}
\affiliation{Department of Physics, Applied Physics, and Astronomy, Binghamton University–SUNY, Binghamton, New York 13902, USA}
\author{Elena R. Margine}
\affiliation{Department of Physics, Applied Physics, and Astronomy, Binghamton University–SUNY, Binghamton, New York 13902, USA}

\begin{abstract}
Superconducting two-dimensional materials, and in particular few‐layer graphene, offer an exciting platform for low‐power electronics, yet the origin of their unconventional superconductivity remains an open question. Prevailing theories, primarily rooted in the Bardeen-Cooper-Schrieffer (BCS) framework that assumes electron-phonon interactions are the main mechanism of superconductivity, struggle to account quantitatively for the observed phenomena. 
Recent studies point to a plasmonic pairing mechanism in graphene systems; however, disentangling the relative contributions of phonon- and plasmon-mediated pairing remains challenging due to the lack of a satisfactory first-principles framework capable of accurately capturing dynamical screening effects in the electronic channel. Here, we present a new theoretical framework that extends the quasiparticle self-consistent GW method to the superconducting phase by coupling it with the Eliashberg treatment of both phonon- and plasmon-mediated interactions. Our approach, termed \red{superconducting quasiparticle GW (s-qpGW)}, is on par with the state-of-the-art Eliashberg theory of superconductivity when applied to bulk metals, and correctly predicts the absence of superconductivity in doped monolayer graphene.
To differentiate s-qpGW from conventional Eliashberg approaches, we study a simple model system, graphene with an artificially enhanced density of states, and demonstrate that s-qpGW captures dynamical Coulomb screening  effects in ways that standard BCS theory cannot. 
 \end{abstract}

\maketitle

\clearpage

\section{Introduction}
Superconducting electronics offer a promising path to low-energy, power-efficient microelectronics, with potential applications in AI and advanced sensors. The 2018 discovery of unconventional superconductivity in graphene superlattices \cite{TBLG18} generated tremendous excitement, and superconductivity has since been observed in commensurate multilayers such as rhombohedral trilayer graphene \cite{TLGexp} as well as in other two-dimensional (2D) layered materials \cite{TMDs1,TMDs2}. 

Two-dimensional layered materials form an ideal playground for exploring unconventional superconductivity as they provide a wealth of tunable parameters such as doping, stacking, relative orientation, strain, and magnetic fields.
Understanding the origin of superconductivity in these systems is crucial for enabling effective materials‐design strategies in next‐generation superconducting electronics. 

However, despite several years of intense research, the mechanisms behind superconductivity in graphene-based materials remain elusive. Early work favored a phonon-mediated pairing mechanism; however, more recent studies \cite{TBLGgeim,ParraMartinez2025} suggest that a plasmon-enabled pairing mechanism is a plausible explanation for superconductivity in graphene superlattices. Current theories—particularly the Bardeen-Cooper-Schrieffer (BCS) framework~\cite{Bardeen1957}—often fall short in explaining the unconventional superconductivity observed in these systems \cite{TBLG18,TLGexp}. For example, while semi-empirical BCS-based models capture general trends of the critical temperature (\(T_c\)) in rhombohedral trilayer graphene as a function of doping and displacement field, they fail to predict the observed superconductivity near the experimentally probed carrier density \cite{SarmasTLG22}. Similarly, recent {\it ab initio} attempts to calculate \(T_c\) in rhombohedral trilayer graphene under a phonon-mediated BCS assumption have overestimated \(T_c\) by about a factor of four \cite{Rubio24}\red{, an error that may be partly attributable to the neglect of quantum fluctuations.} These limitations underscore the need for a more accurate computational framework to guide practical applications.

In 2D systems, plasmons are acoustic (vanishing in energy at long wavelengths \cite{SpataruBLG23}) and may directly compete with phonons in establishing superconductivity. Studying the balance between phonon‐ and plasmon‐mediated pairing mechanisms is thus very important for understanding the origin of superconductivity in 2D materials and in graphene systems in particular~\cite{Takada1978,Akashi2013,Akashi2014,Sanna2020,Davydov2020,Akashi2022,Veld2023}.
To this end, we propose a newly developed theoretical framework, which incorporates the phonon-based pairing mechanism within the standard, state-of-the-art Eliashberg theory of superconductivity~\cite{Eliashberg1960,Eliashberg1961}, while extending beyond BCS to include plasmonic pairing effects. Importantly, this method addresses an outstanding issue related to self-consistency and the lack of vertex corrections in the electronic channel \cite{Davydov2020,Dassarma2025}.

In \sref{Formalism} we present our extended Migdal-Eliashberg formalism, introducing several approaches for treating the Coulomb channel. 
In \sref{Results} we benchmark these approaches on two prototypical cases—doped monolayer graphene (2D) and bulk Nb (3D)—and on a simple 2D model. 
Finally, in \sref{Conclusion} we summarize our findings and discuss the outlook for materials with competing phonon- and plasmon-mediated pairing channels.

\section{Eliashberg theory of Superconductivity}
\label{Formalism}

Within the Matsubara-frequency Green’s-function formalism, Migdal-Eliashberg theory casts the Dyson equation \red{in a reduced \(2\times2\) Nambu space for pairing between time-reversed states~\cite{Gorkov1958,Nambu1960,Margine2013,Lane2026,Spataru2021} as}
\begin{equation}
[\hat{G}_{n\mathbf{k}}(i\omega_{j})]^{-1}
= [\hat{G}^0_{n\mathbf{k}}(i\omega_{j})]^{-1}
  - \hat{\Sigma}_{n\mathbf{k}}(i\omega_{j})\,,
\label{eqG}  
\end{equation}
where $\hat{G}^0_{n\mathbf{k}}(i\omega_{j})$ is the Green's function for the system in the normal state, and within a noninteracting or mean-field description such as
density functional theory (DFT):
\begin{equation}
[\hat{G}^0_{n\mathbf{k}}(i\omega_{j})]^{-1}
= i\,\omega_{j}\,\hat{\tau}_{0}
  - \epsilon_{n\mathbf{k}}\,\hat{\tau}_{3}\,.
\end{equation}
Here $\hat{\tau}_{i}$ ($i=0,1,2,3$) denote the Pauli matrices and
$\epsilon_{n\mathbf{k}}$ represents the energy (measured with respect to the
Fermi level $E_F$) of a Bloch state with momentum $\mathbf{k}$ and band
index $n$ (throughout this work we use the band-diagonal approximation, that
is, no band mixing). We note that in the case of DFT-based {\it ab initio} approaches - as in our work - where the Bloch state is the Kohn-Sham state, one needs to subtract the contribution of the exchange-correlation potential $ v^{xc}$, $\hat{\Sigma} \rightarrow \hat{\Sigma}-v_{xc}\,\hat{\tau}_{3}$ in \eref{eqG} \cite{Davydov2020}.

The electron self‐energy $\hat\Sigma_{n\mathbf{k}}(i\omega_{j})$ is evaluated
to leading (first) order in the screened Coulomb interaction
\beq
  W = W^{ph} + W^{C}\,,
\eeq
where $W^{ph}$ and $W^{C}$ are the phonon‐mediated and purely electronic
(Coulomb) contributions, respectively. The electronic screening is treated within
the random phase approximation (RPA),
\beq
  W^{C} = \epsilon^{-1} v,
  \quad
  \epsilon = 1 - v\,P^{\rm RPA},
\eeq
with $v$ the bare Coulomb potential and $P^{\rm RPA}$ the RPA irreducible polarizability.
As first derived by Hedin and Lundqvist \cite{Hedin1970}, the phonon part can be written
\beq
  W^{ph} = W^{C} D W^{C},
  \label{Wph-D}
\eeq
where $D$ is the phonon propagator~\cite{Giustino2017}.

The self‐energy is then written
\begin{equation}
\begin{split}
\hat\Sigma_{n\mathbf{k}}(i\omega_j)
&= -\,T\sum_{n'\mathbf{k}' j'}
   \hat{\tau}_{3}\,\hat G_{n'\mathbf{k}'}(i\omega_{j'})\,\hat{\tau}_{3} \\
&\quad\times
   W_{n\mathbf{k},n'\mathbf{k}'}(i \omega_j - i \omega_{j'})\,.
\end{split}
\label{Migdal-GW}
\end{equation}
\red {Here, \(T\) denotes the temperature. The sums over \(n'\mathbf{k}'\) in Eq.~\eqref{Migdal-GW} include all states lying within an energy window typically several times larger than the relevant phonon energies, consistent with the EPW implementation of the anisotropic Eliashberg equations \cite{Margine2013,Ponce2016,Lee2023}.}

Equation~\eqref{Migdal-GW} incorporates two approximations: i) Migdal’s approximation~\cite{Migdal1958} for the
electron-phonon coupling (the Fan-Migdal electron-phonon self-energy), and
ii) the GW approximation \cite{Hedin1965,Hybertsen1986,Strinati1980,Strinati1982,Benedict2002} for the electron self-energy. Both approximations
neglect vertex corrections.

Next, we use the following notation for the Nambu components of
$\hat{\Sigma}_{n\mathbf{k}}(i\omega_j)$:
\beq
\hat{\Sigma}_{n\mathbf{k}}(i \omega_j) =
\begin{pmatrix}
[\Sigma_{n\mathbf{k}}(i\omega_j)]_{11} & \phi_{n\mathbf{k}}(i\omega_j) \\
\phi_{n\mathbf{k}}(i\omega_j)         & [\Sigma_{n\mathbf{k}}(i\omega_j)]_{22}
\end{pmatrix},
\eeq
where the off-diagonal components $\phi_{n\mathbf{k}}$ denote the anomalous self-energy, {\it i.e.}, the superconducting order-parameter.

The normal (diagonal) components of $\hat{\Sigma}_{n\mathbf{k}}(i\omega_j)$ can be written as the sum of odd and even (in $i\omega_j$ ) scalar contributions and take the form
\beq
[\Sigma_{n\mathbf{k}}(i\omega_j)]_{11}
= i\omega_j \bigl[1-Z_{n\mathbf{k}}(i\omega_j)\bigr]
  + \chi_{n\mathbf{k}}(i\omega_j),
\label{Sigma_11_Z_chi}
\eeq
where
%\red{[RM: removed "-" on the LHS of (9).]}
\beq
i\omega_j \bigl[1-Z_{n\mathbf{k}}(i\omega_j)\bigr]
\equiv \frac{1}{2}\Big(
        [\Sigma_{n\mathbf{k}}(i\omega_j)]_{11}
      - [\Sigma_{n\mathbf{k}}(-i\omega_j)]_{11}
      \Big),
\eeq
and
\beq
\chi_{n\mathbf{k}}(i\omega_j)
\equiv \frac{1}{2}\Big(
        [\Sigma_{n\mathbf{k}}(i\omega_j)]_{11}
      + [\Sigma_{n\mathbf{k}}(-i\omega_j)]_{11}
      \Big),
\eeq
where both $Z_{n\mathbf{k}}(i\omega_j)$ and $\chi_{n\mathbf{k}}(i\omega_j)$
are even functions of $i\omega_j$ and $\mathbf{k}$.
Similarly, using that
$[\Sigma_{n\mathbf{k}}(i\omega_j)]_{22}
 = -[\Sigma_{n\mathbf{k}}(-i\omega_j)]_{11}$, it follows that
\beq
[\Sigma_{n\mathbf{k}}(i\omega_j)]_{22}
  = i\omega_j \bigl[1 - Z_{n\mathbf{k}}(i\omega_j)\bigr]
    - \chi_{n\mathbf{k}}(i\omega_j).
\label{Sigma_22_Z_chi}
\eeq

The even contribution $\chi_{n\mathbf{k}}(i\omega_j)$ mostly produces a static
shift of the quasiparticle (QP) energies,
$\epsilon_{n\mathbf{k}}\rightarrow
       \epsilon_{n\mathbf{k}}+\chi_{n\mathbf{k}}(i\omega_j)-v^{xc}_{n\mathbf{k}}$. 
In practice one neglects this contribution, 
$\chi_{n\mathbf{k}}(i\omega_j)-v^{xc}_{n\mathbf{k}}\rightarrow 0$, since in a metallic system QP energies near the Fermi level differ only slightly from their Kohn-Sham values.

The odd contribution $i\omega_j[1-Z_{n\mathbf{k}}(i\omega_j)]$
%\red{[RM: removed "-"]} 
is parametrized by the mass renormalization function
$Z_{n\mathbf{k}}(i\omega_j)$, which can be interpreted as a renormalization of the effective mass of the Cooper pairs.
%[not to be confused with the QP renormalization constant
%$Z^{QP}_{n\mathbf{k}}(\omega)=
%      [1-d\Re\,[\Sigma_{n\mathbf{k}}(\omega)]_{11}/d\omega]^{-1}_{|\omega=E^{QP}}$],
%and it can also renormalize the QP energies.

It is difficult to decide {\it a priori} whether to neglect or not the Coulomb
contribution $Z^{C}_{n\mathbf{k}}(i\omega_j)$ to the mass renormalization
function $Z_{n\mathbf{k}}(i\omega_j)$. However, {\it a posteriori} one finds
that retaining $Z^{C}_{n\mathbf{k}}(i\omega_j)$ partially mitigates (though
does not fully eliminate) the artificial overestimation of $T_{c}$ produced by
full dynamical Coulomb effects in the superconducting self-consistent GW (s-GW)
scheme (see Sec.~\ref{sec:scGW}) \cite{Davydov2020}. Accordingly, we keep
$Z^{C}_{n\mathbf{k}}(i\omega_j)$ during the s-GW calculations presented here.

Neglecting the energy shift $\chi_{n\mathbf{k}}(i\omega_j)$, the
equations for $\phi_{n\mathbf{k}}(i\omega_j)$ and $Z_{n\mathbf{k}}(i\omega_j)$
can be written as
\beq
\begin{split}
\phi_{n\mathbf{k}}(i\omega_j)
&= - T \sum_{n'\mathbf{k}' j'}
   [G_{n'\mathbf{k}'}(i\omega_{j'})]_{12} \\
&\quad\times
   W_{n\mathbf{k},n'\mathbf{k}'}(i\omega_j-i\omega_{j'}),
\end{split}
\label{Sigma_12}
\eeq
\beq
\begin{split}
i\omega_j \bigl[1-Z_{n\mathbf{k}}(i\omega_j)\bigr]
&= - T \sum_{n'\mathbf{k}' j'}
      [G_{n'\mathbf{k}'}(i\omega_{j'})]_{11} \\
&\times
   W_{n\mathbf{k},n'\mathbf{k}'}(i\omega_j-i\omega_{j'}),
\end{split}
\label{Sigma_11}
\eeq
or
\begin{equation}
\begin{split}
\phi_{n\mathbf{k}}(i\omega_j)
&= T \sum_{n'\mathbf{k}' j'}
       \frac{\phi_{n'\mathbf{k}'}(i\omega_{j'})}
            {\det\hat{G}^{-1}_{n'\mathbf{k}'}(i\omega_{j'})} \\
&\quad\times
   W_{n\mathbf{k},n'\mathbf{k}'}(i\omega_j - i\omega_{j'}),
\end{split}
\label{phi_def}
\end{equation}
\begin{equation}
\begin{split}
Z_{n\mathbf{k}}(i\omega_j)
&= 1+\frac{T}{\omega_j}
   \sum_{n'\mathbf{k}' j'}
   \frac{\omega_{j'}\,Z_{n'\mathbf{k}'}(i\omega_{j'})}
        {\det\hat{G}^{-1}_{n'\mathbf{k}'}(i\omega_{j'})} \\
&\quad\times
   W_{n\mathbf{k},n'\mathbf{k}'}(i\omega_j - i\omega_{j'}),
\end{split}
\label{Z_def}
\end{equation}
where
\begin{equation}
\begin{split}
\det\hat{G}^{-1}_{n\mathbf{k}}(i\omega_{j}) 
= -\left[\omega_{j}^2\,Z^{2}_{n\mathbf{k}}(i\omega_{j})
 + \epsilon_{n\mathbf{k}}^{2}
 + \phi^{2}_{n\mathbf{k}}(i\omega_{j})\right].
\end{split}
\label{eq:Theta}
\end{equation}

\red{
Following standard Eliashberg notation \cite{Margine2013}, the superconducting gap function is defined in terms of the anomalous self-energy and the mass-renormalization function as
\begin{equation}
\Delta_{n\mathbf{k}}(i\omega_j)
=
\frac{\phi_{n\mathbf{k}}(i\omega_j)}{Z_{n\mathbf{k}}(i\omega_j)}.
\label{gap_def}
\end{equation}
}

Finally, we decompose $\phi_{n\mathbf{k}}(i\omega_{j})$ and
$Z_{n\mathbf{k}}(i\omega_{j})$ into separate phonon and Coulomb contributions
through
\beq
\phi_{n\mathbf{k}}(i\omega_{j})
\equiv \phi^{ph}_{n\mathbf{k}}(i\omega_{j})
     + \phi^{C}_{n\mathbf{k}}(i\omega_{j}),
\eeq
and
\beq
Z_{n\mathbf{k}}(i\omega_{j})
\equiv 1 + Z^{ph}_{n\mathbf{k}}(i\omega_{j})
         + Z^{C}_{n\mathbf{k}}(i\omega_{j}),
\eeq
with the individual contributions being defined according to Eqs.~\eqref{phi_def} and \eqref{Z_def}, where on the right-hand side we replace $W$ by $W^{ph}$
or $W^{C}$ as appropriate:
\begin{equation}
\begin{split}
\phi^{ph/C}_{n\mathbf{k}}(i\omega_j)
&= T \sum_{n'\mathbf{k}' j'}
       \frac{\phi_{n'\mathbf{k}'}(i\omega_{j'})}
            {\det\hat{G}^{-1}_{n'\mathbf{k}'}(i\omega_{j'})} \\
&\quad\times
   W^{ph/C}_{n\mathbf{k},n'\mathbf{k}'}(i\omega_j - i\omega_{j'}),
\end{split}
\label{phi_def2}
\end{equation}
\begin{equation}
\begin{split}
Z^{ph/C}_{n\mathbf{k}}(i\omega_j)
&= \frac{T}{\omega_j}
   \sum_{n'\mathbf{k}' j'}
   \frac{\omega_{j'}\,Z_{n'\mathbf{k}'}(i\omega_{j'})}
        {\det\hat{G}^{-1}_{n'\mathbf{k}'}(i\omega_{j'})} \\
&\quad\times
   W^{ph/C}_{n\mathbf{k},n'\mathbf{k}'}(i\omega_j - i\omega_{j'}).
\end{split}
\label{Z_def2}
\end{equation}

In the next section we turn to several strategies for the electronic channel, in all cases treating the phonon contribution on an equal footing; see Appendix A (\sref{AppendixA}) for the detailed phonon-channel equations. The superconducting extensions in Nambu space of the normal‐state fully self‐consistent GW and of the quasiparticle self‐consistent GW are denoted s‐GW and s‐qpGW, respectively; we also introduce the static variant s‐GW$_{\text{static}}$. Throughout this paper, the prefix “s-” denotes “superconducting”, and in all three acronyms the letter “W” refers to the purely electronic screened Coulomb interaction W$^C$.

\subsection{The fully self-consistent GW approach}
\label{sec:scGW}

Without any further approximation the Coulomb contributions can be written as 
\begin{equation}
\begin{split}
\phi^{C}_{n\mathbf{k}}(i\omega_{j})
&= {T}
   \sum_{n'\mathbf{k}' j'}
   \frac{\phi_{n'\mathbf{k}'}(i\omega_{j'})}
        {\det\hat{G}^{-1}_{n'\mathbf{k}'}(i\omega_{j'})} \\
&\quad\times     
   \,W^{C}_{n\mathbf{k},n'\mathbf{k}'}(i\omega_j-i\omega_{j'}) ,
\end{split}
\label{scGW_phiC}
\end{equation}
where the negative denominator on the right-hand side of \eref{scGW_phiC} ensures that the Coulomb term promotes superconductivity only when the effective Coulomb screening is attractive, that is when $W^C<0$. 

Finally, the Coulomb contribution to the mass renormalization function reads
\begin{equation}
\begin{split}
Z^{C}_{n\mathbf{k}}(i\omega_{j})
&= \frac{T}{\omega_{j}}
   \sum_{n'\mathbf{k}' j'}
   \frac{\omega_{j'}\,Z_{n'\mathbf{k}'}(i\omega_{j'})}
        {\det\hat{G}^{-1}_{n'\mathbf{k}'}(i\omega_{j'})} \\
&\quad\times      
   \,W^{C}_{n\mathbf{k},n'\mathbf{k}'}(i\omega_j-i\omega_{j'}) .
\end{split}
\label{scGW_ZC}
\end{equation}
Equations \eqref{scGW_phiC} and \eqref{scGW_ZC} represent the fully frequency-dependent,
self-consistent GW treatment of the electronic channel within the Eliashberg formalism, which we refer to as ``s-GW''. 

\subsection{The standard (static) Eliashberg approach}
\label{subsec:BCS}

The state-of-the-art,
standard Eliashberg approach employs the static approximation for the screened
Coulomb interaction,
$W^{C}_{n\mathbf{k},n'\mathbf{k}'}(i\omega_j)
 \rightarrow W^{C}_{n\mathbf{k},n'\mathbf{k}'}(0)$.
This approximation neglects the high-energy scale relevant to chemical bonding
and by construction excludes any plasmonic-pairing effects. This is
justified in bulk metals where electron excitations (that is plasmons) are
significantly higher in energy than phonons and should not play an important
role in the formation of Cooper pairs. However, the approximation may be questioned in 2D systems where acoustic plasmons compete with 
phonons and dynamical screening effects and/or plasmonic pairing may become important.

Within the static approximation, \eref{scGW_phiC} yields a frequency-independent Coulomb contribution to the pairing function
\begin{equation}
\begin{split}
\phi^{C,{\rm stat}}_{n\mathbf{k}}
&= {T}
   \sum_{n'\mathbf{k}' j'}
   \frac{\phi_{n'\mathbf{k}'}(i\omega_{j'})}
        {\det\hat{G}^{-1}_{n'\mathbf{k}'}(i\omega_{j'})}
   \,W^{C}_{n\mathbf{k},n'\mathbf{k}'}(0) .
\end{split}
\label{BCS_static}
\end{equation}
$Z^{C,{\rm stat}}_{n\mathbf{k}}(i\omega_{j})=0$ by virtue of the even
parity of $Z_{n\mathbf{k}}(i\omega_{j})$ and
$\phi_{n\mathbf{k}}(i\omega_{j})$ with respect to the imaginary Matsubara frequency
$i \omega_j$, which makes the summand in \eref{scGW_ZC} an odd function
of $i \omega_j$ and thus identically zero upon summation. We denote the electrostatic treatment of the Coulomb interaction by  ``s-GW$_{static}$''.

\subsection{The semi-empirical $\mu^*$ approach}
\label{subsec:mu_star}

Due to the significant computational difficulty in dealing with the electronic
channel within the Eliashberg approach, a popular practice is to treat the Coulomb
term \eref{BCS_static} in a semi-empirical manner~\cite{Morel1962}. The factor
$N_F\,W^{C}_{n\mathbf{k},n'\mathbf{k}'}(0)$ is replaced by a dimensionless
constant
$\mu_c = N_F\,\langle W^{C}_{n\mathbf{k},n'\mathbf{k}'}(0)\rangle$, where
$\langle \cdots \rangle$ represents a double Fermi-surface average over
$\mathbf{k}$ and $\mathbf{k}'$, and $N_F$ is the density of states per spin at the Fermi level.
Coupled with an additional cutoff in the
Matsubara frequency $\omega_{cut}^{ph}$ (chosen large enough to ensure
$\phi^{ph}_{n\mathbf{k}}(i\omega_{cut}^{ph})\approx 0$, for example
$\omega_{cut}^{ph} \approx 1$ eV), the pairing function simplifies into a frequency-
and momentum-independent parameter
\begin{equation}
\begin{split}
\phi^{C}_{\mu^*}
&= \frac{T}{N_{F}}
   \sum_{n'\mathbf{k}' j'}^{\omega_{j'}<\omega_{cut}^{ph}}
   \frac{\phi_{n'\mathbf{k}'}(i\omega_{j'})}
        {\det\hat{G}^{-1}_{n'\mathbf{k}'}(i\omega_{j'})}\,\mu^*_c ,
\end{split}
\label{mu_star}
\end{equation}
with
\beq
\mu^*_{c} = \frac{\mu_{c}}{1 + \mu_{c} \ln\left({E_{F}}/{\omega_{cut}^{ph}}\right)}.
\eeq

The parameter $\mu^*_{c}$ is an approximate measure of the Coulomb repulsion
term~\cite{Morel1962} and it has been tabulated for a large variety of bulk superconductors,
with values in the range $0.1 - 0.2$~\cite{Allen1983,Pellegrini2024}.

\subsection{The quasiparticle self-consistent GW approach}
\label{subsec:s-qpGW}

\red{
In this work we introduce a new strategy in which the frequency dependence of the Coulomb self-energy \(\hat{\Sigma}^C\) is eliminated by a symmetrized convolution with the QP spectral function \(\hat{A}\). The essential steps of our extension to the superconducting phase are summarized here, while \appref{AppendixB} provides the full derivation.

The QP spectral function is defined in terms of the retarded and advanced full Green's functions \(\hat{G}^{r,a}_{n\mathbf{k}}(\omega)\), obtained by analytic continuation from the Matsubara imaginary axis to the real axis according to the prescription \(i\omega \to \omega \pm i\eta\):
\[
\hat{A}_{n\mathbf{k}}(\omega)=
\frac{\hat{G}^{a}_{n\mathbf{k}}(\omega)-\hat{G}^{r}_{n\mathbf{k}}(\omega)}{2\pi i}.
\]

The central step of this strategy is the definition of a static Coulomb self-energy through the symmetrized convolution
\begin{equation}
\begin{split}
\hat{\tilde{\Sigma}}^{C}_{n\mathbf{k}}
=
\frac{1}{2}\int d\omega \,
\Re \hat{\Sigma}^{C}_{n\mathbf{k}}(\omega)\hat{A}_{n\mathbf{k}}(\omega)
\\
+
\frac{1}{2}\int d\omega \,
\hat{A}_{n\mathbf{k}}(\omega)\Re \hat{\Sigma}^{C}_{n\mathbf{k}}(\omega) .
\end{split}
\end{equation}
In the normal-state limit, this construction reduces to the quasiparticle self-consistent GW method \cite{Faleev2004,vanSchilfgaarde2006}. 

Within Eliashberg theory, this construction makes the Coulomb channel contribute only through a frequency-independent anomalous self-energy (with no corresponding contribution to the mass-renormalization function \(Z\)):
\begin{equation}
\tilde{\phi}^{C}_{n\mathbf{k}}
=
\int d\omega \,
\Re \phi^{C}_{n\mathbf{k}}(\omega)\,[\hat{A}_{n\mathbf{k}}(\omega)]_{11}
\approx
\Re \phi^{C}_{n\mathbf{k}}(E^{\mathrm{QP}}_{n\mathbf{k}}),
\label{phi_tilde_main}
\end{equation}
where \([\hat{A}_{n\mathbf{k}}(\omega)]_{11}\) is the normal component of the QP spectral function.

Equation~(\ref{phi_tilde_main}) extends the QP approximation within GW to the superconducting phase. In the normal state, the QP approximation is known to partially compensate for the effects of the missing vertex corrections in fully self-consistent GW \cite{Holm1998,Bruneval2006,Shishkin2007}, restoring band gaps and QP energies to within a few tenths of an eV of experiment \cite{vanSchilfgaarde2006}.
Applying the QP approximation to the anomalous self-energy yields the superconducting quasiparticle GW or ``s-qpGW'' scheme: a practical, well-behaved \emph{ab initio} framework for dynamical Coulomb screening in superconductors.

We emphasize that within s-qpGW the electron-plasmon coupling still enters through the full energy dependence of \(W^C\), which appears in \eref{scGW_phiC}. Indeed, at each iteration of the self-consistent cycle we evaluate \(\phi^C_{n\mathbf{k}}(\omega)\) by applying Pad\'e-approximant analytic continuation~\cite{Vidberg77}, as implemented in a subroutine of the EPW code \cite{Margine2013,Ponce2016,Lee2023}, to its Matsubara-axis counterpart in \eref{scGW_phiC}. In future developments, it may be advantageous to evaluate the Coulomb self-energy directly on the real-frequency axis rather than obtain it by analytic continuation from the Matsubara axis.
}

\section{Results}
\label{Results}

We consider two prototypical cases, with different dimensionalities, where dynamical screening plays very different roles: (i) doped monolayer graphene in 2D, which hosts acoustic plasmons that vanish at long wavelength, and (ii) bulk Nb in 3D, whose optical plasmons lie several eV above the phonon spectrum.  This contrast provides an ideal testbed for our extended Eliashberg-GW framework, in particular s-GW and s-qpGW. \red{All computational details are provided in Sec.~\ref{Methods}. \tref{table_1} summarizes the superconducting critical temperatures obtained for bulk Nb and doped monolayer graphene within the different Coulomb-channel approximations considered below.}

\begin{table}[t]
\caption{
\red{Summary of the superconducting critical temperature \(T_c\) (in K) obtained for bulk Nb and doped monolayer graphene within the different Coulomb-channel approximations. When a superconducting solution is not found, we report ``n/a''. For Nb, spin-fluctuation contributions are not included.}
}
\label{tab:summary}
\begin{ruledtabular}
\begin{tabular}{lccc}
 & s-GW & s-GW$_{\mathrm{static}}$ & s-qpGW \\
\hline
Bulk Nb & 23.5 & 13.5 & 14.0 \\
Doped graphene & \(> 3\) & n/a & n/a \\
\end{tabular}
\end{ruledtabular}
\label{table_1}
\end{table}

\begin{figure}
\includegraphics[trim=-0 0 -0 0,clip,width=\columnwidth]{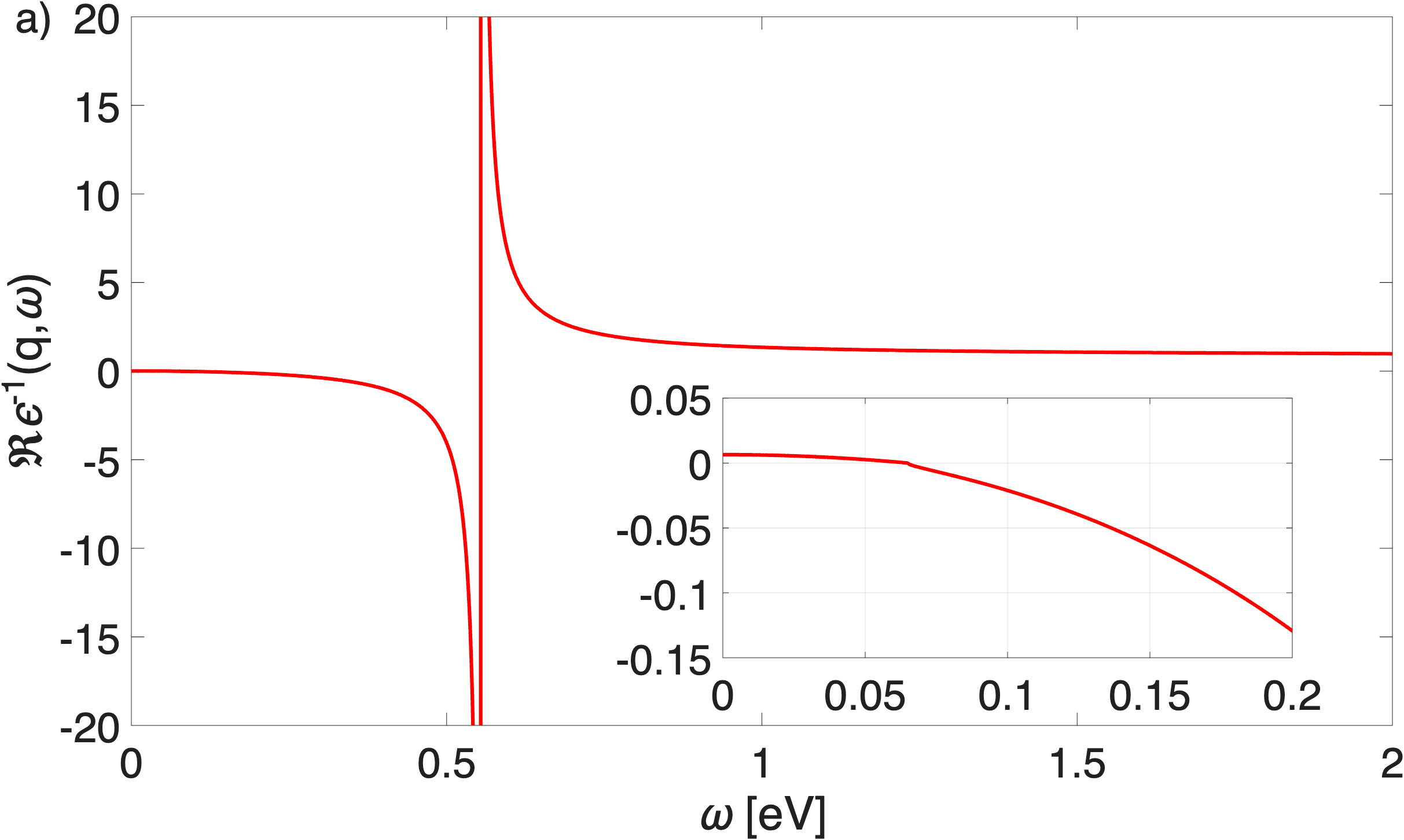}
\includegraphics[trim=-0 0 -0 0,clip,width=\columnwidth]{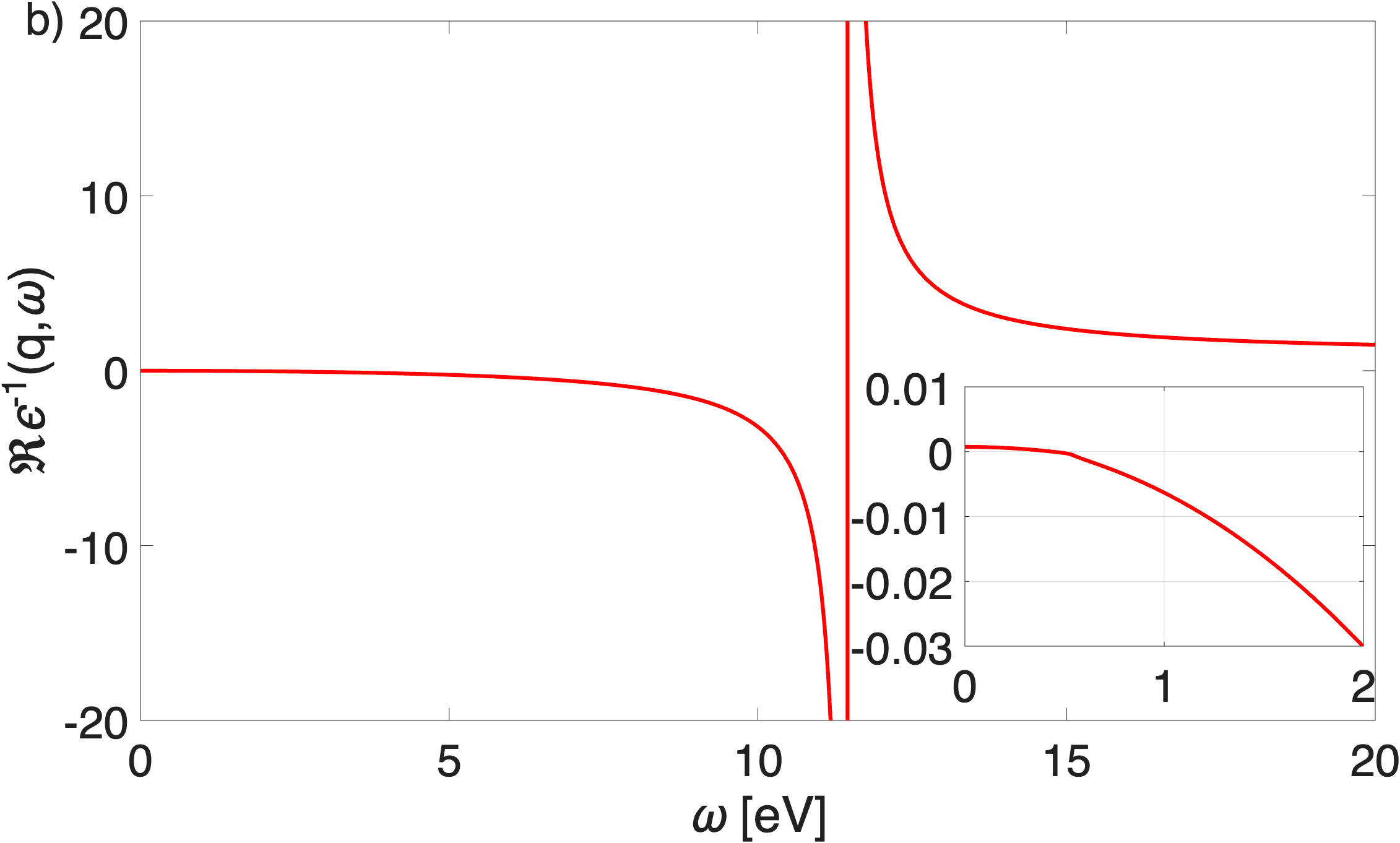}
\caption{a) Real part of the inverse dielectric function of doped graphene for momentum \(q = 10^{\red{8}}\,\mathrm{m}^{-1}\). b) Real part of the inverse dielectric function of bulk Nb for momentum \(q = 10^{\red{9}}\,\mathrm{m}^{-1}\). The inset in each panel shows an enlarged view near \(\omega = 0\), where \(\Re\,\epsilon^{-1}(\omega)\) changes from positive to negative.}
\label{Wel_gr}
\end{figure}

Figure~\ref{Wel_gr}a shows the dynamical response, through the real part of the inverse dielectric function \(\epsilon^{-1}(q,\omega)\), for doped graphene, for a momentum \(q=10^{8}\,\mathrm{m}^{-1}\). We consider an $n$-type doping level relatively large but below the capacitive limit of graphene \cite{Ye2011}, namely $n=9.5\times 10^{13} \text{/cm}^2$. This corresponds to a Fermi level positioned $\approx 1$ eV above the Dirac point, accommodating $0.05$ extra \red{electrons} per unit cell. 
\red{ We use the analytical dielectric function of Ref.~\cite{HwangDasSarma2007}, which is based on an effective low-energy Hamiltonian and therefore does not include higher-energy bands that could generate additional optical-like plasmon branches.
The energy of the intraband plasmon branch vanishes as \(q\to0\) and increases rapidly with momentum away from the long-wavelength limit. 
For the momentum used in \fref{Wel_gr}a, the intraband plasmon energy is about \(0.6\) eV, as seen from the rapid variation in \(\mathrm{Re}\,\epsilon^{-1}(q,\omega)\).}

In general, the electronic Coulomb interaction $W^C(\omega)$ is repulsive in the electrostatic limit $\omega=0$, but turns attractive as $\omega$ approaches the plasmon energy $\omega_{pl}$.  This is especially pronounced in 2D systems (e.g.\ doped graphene), as clearly seen in the inset of Fig.~\ref{Wel_gr}a).  By contrast, in bulk metals plasmons are optical modes sitting several eV above the phonon spectrum, and thus do not overlap with low‐energy phonons. 
\red{This can be seen in Fig.~\ref{Wel_gr}b, which shows the real part of the inverse dielectric function of bulk Nb, obtained within the Lindhard screening model \cite{Lindhard_note} for momentum \(q = 10^{9}\,\mathrm{m}^{-1}\).}
As a result, we expect dynamical screening (i.e.\ the frequency-dependent Coulomb response) to play a far more significant role in pairing for 2D graphene than in 3D metals.

\red{We note that our choice for analytical dielectric models  is motivated mainly by the computational difficulty of treating dynamical screening explicitly in the electronic channel. We expect the analytical form to be reliable in the small-\(q\) regime relevant to the low-energy plasmon response, where local-field effects are generally weak. While a fully \textit{ab initio} treatment of the frequency-dependent dielectric response may be computationally very demanding within the Eliashberg framework, a useful strategy for future superconductivity calculations could be to compute the dielectric response from first principles and then describe the plasmonic contribution to screening by fitting to an effective low-energy model \cite{Caruso2016}.
}

\begin{figure}
\includegraphics[trim=0 0 0 0,clip,width=\columnwidth]{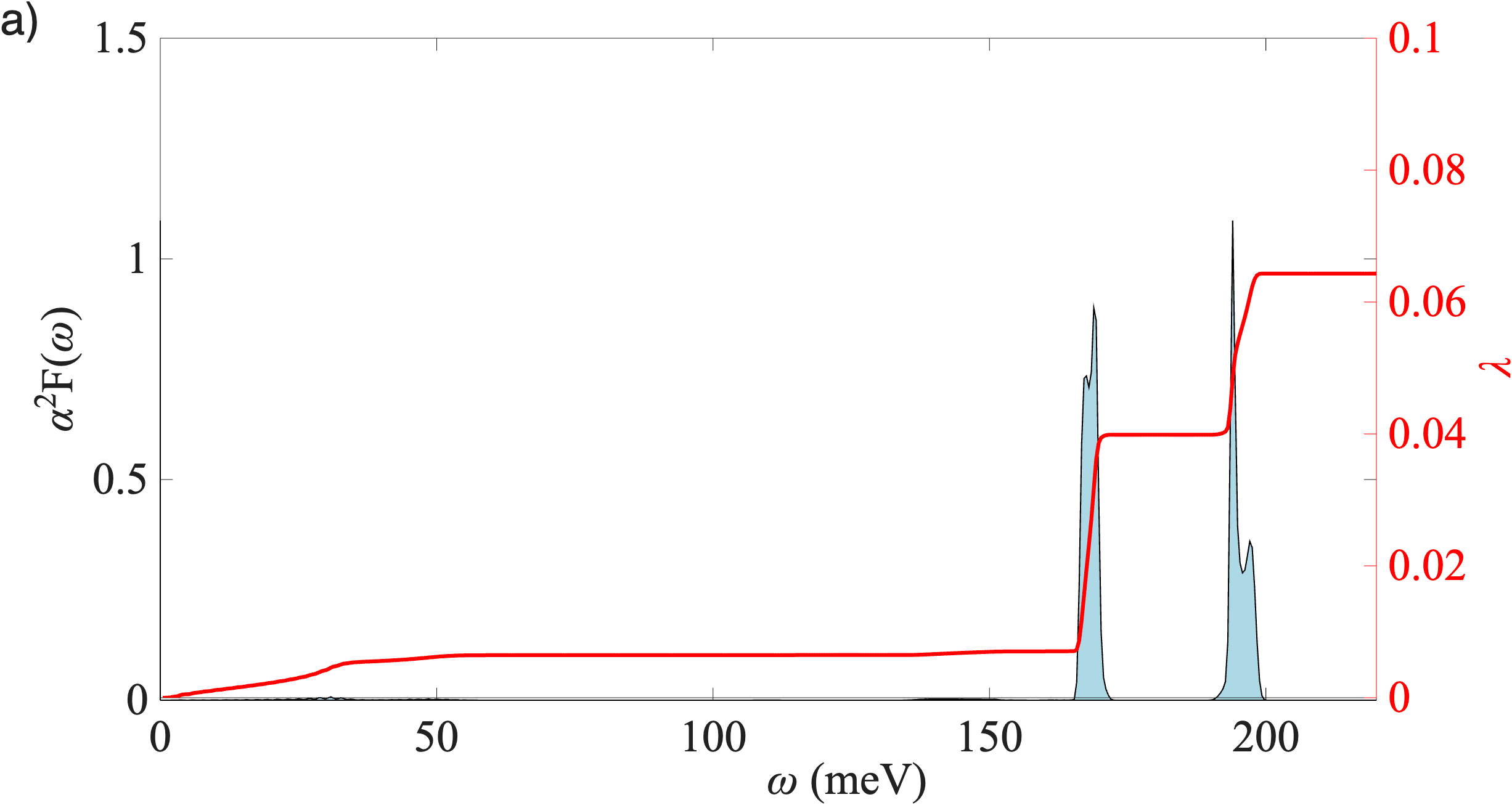}
\includegraphics[trim=0 0 0 0,clip,width=\columnwidth]{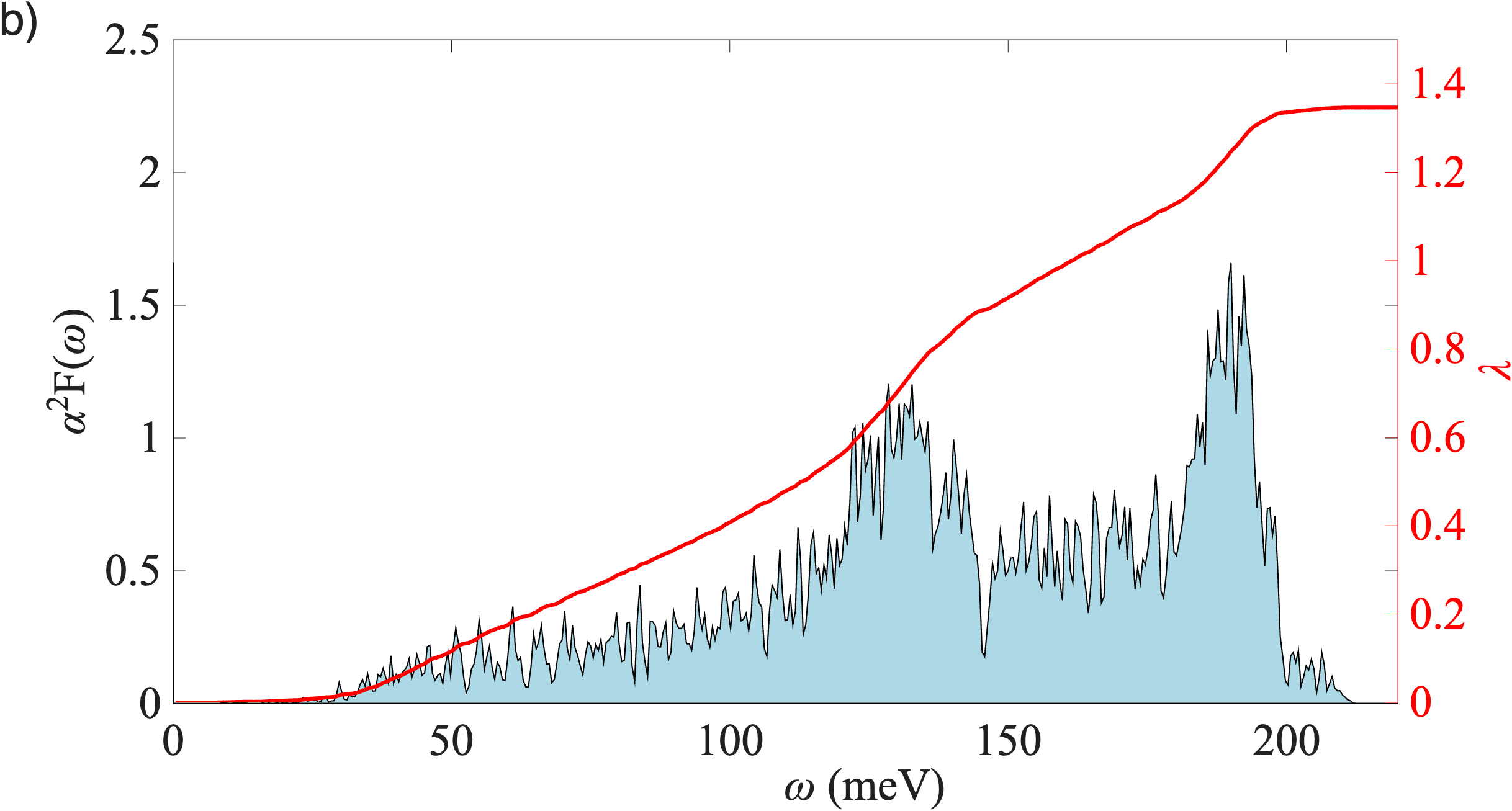}
\caption{Calculated isotropic Eliashberg spectral
function $\alpha^2F$ (black line), and cumulative contribution to
the electron-phonon coupling strength $\lambda$ (red line) for:  
a) Doped monolayer graphene, and b) bulk Nb.}
\label{a2f_Gr_Nb}
\end{figure}

We compute the superconducting properties of doped monolayer graphene and bulk Nb from first principles by means of the fully anisotropic Migdal-Eliashberg formalism implemented in the EPW code \cite{Ponce2016,Lee2023,Margine2013}. Figure~\ref{a2f_Gr_Nb} shows the calculated isotropic Eliashberg spectral function $\alpha^2F(\omega)$ and its cumulative coupling $\lambda(\omega)$ (definitions are given in \appref{AppendixA}) for (a) doped graphene and (b) bulk Nb. The two spectra exhibit very different phononic pairing landscapes, reflecting the highly selective and rather weak coupling in doped graphene versus the broad, strong coupling in Nb. In $n$-doped graphene, where the Fermi surface consists of small pockets around the Dirac point, the restricted phase space favors scattering by a narrow set of high-energy optical phonons~\cite{Margine2014}. In Nb, the large multi-sheet 3D Fermi surface provides ample phase space so that electrons couple efficiently to phonons across essentially the entire spectrum~\cite{Mori2024}.

With these contrasting dynamical screening and electron-phonon behaviors in hand, we now present our first-principles s-GW and s-qpGW results for doped graphene and bulk Nb.

\subsection{s-GW applied to doped graphene and Nb}

\begin{figure}
\includegraphics[trim=-0 0 -0 0,clip,width=\columnwidth]{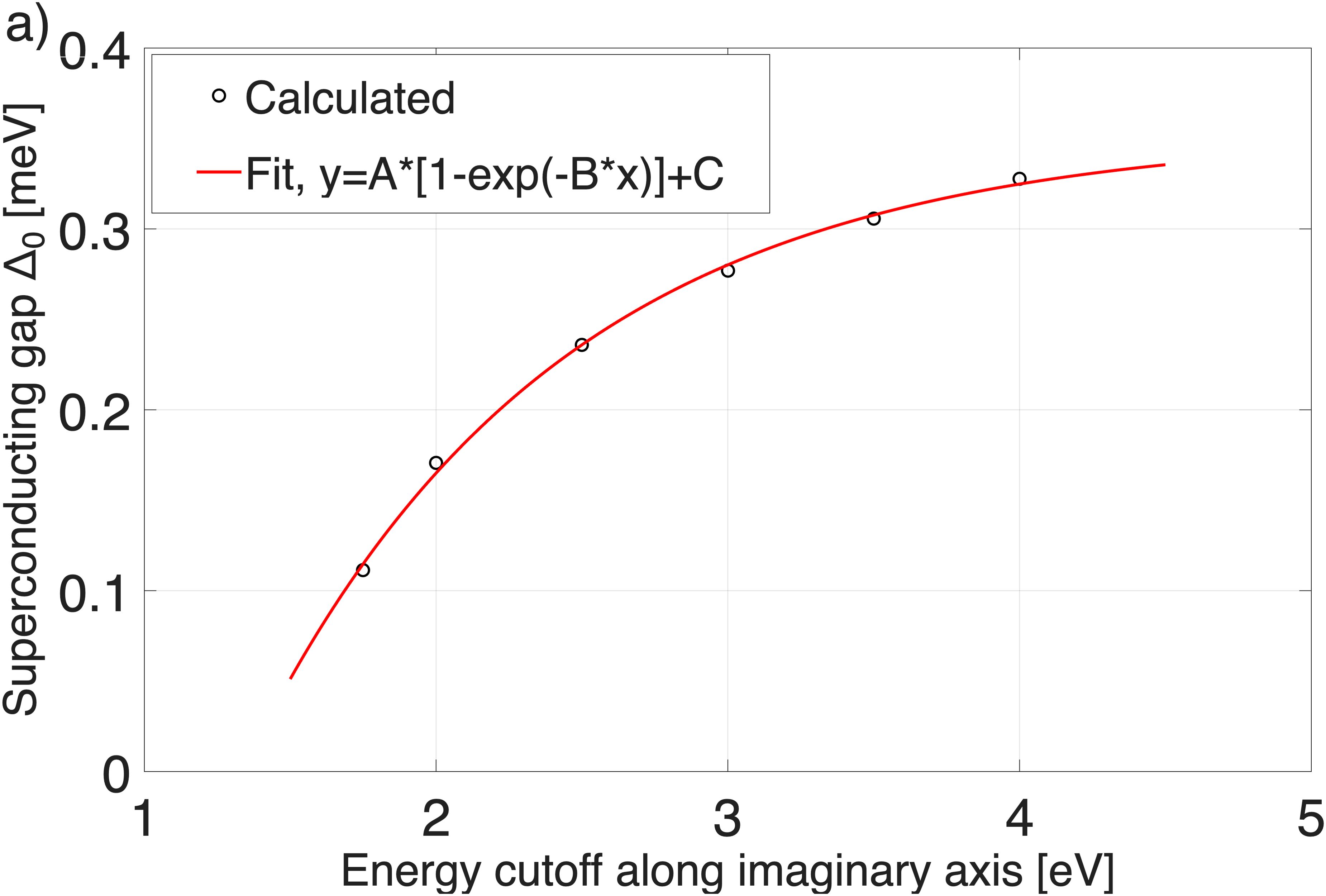}
\includegraphics[trim=-0 0 -0 0,clip,width=\columnwidth]{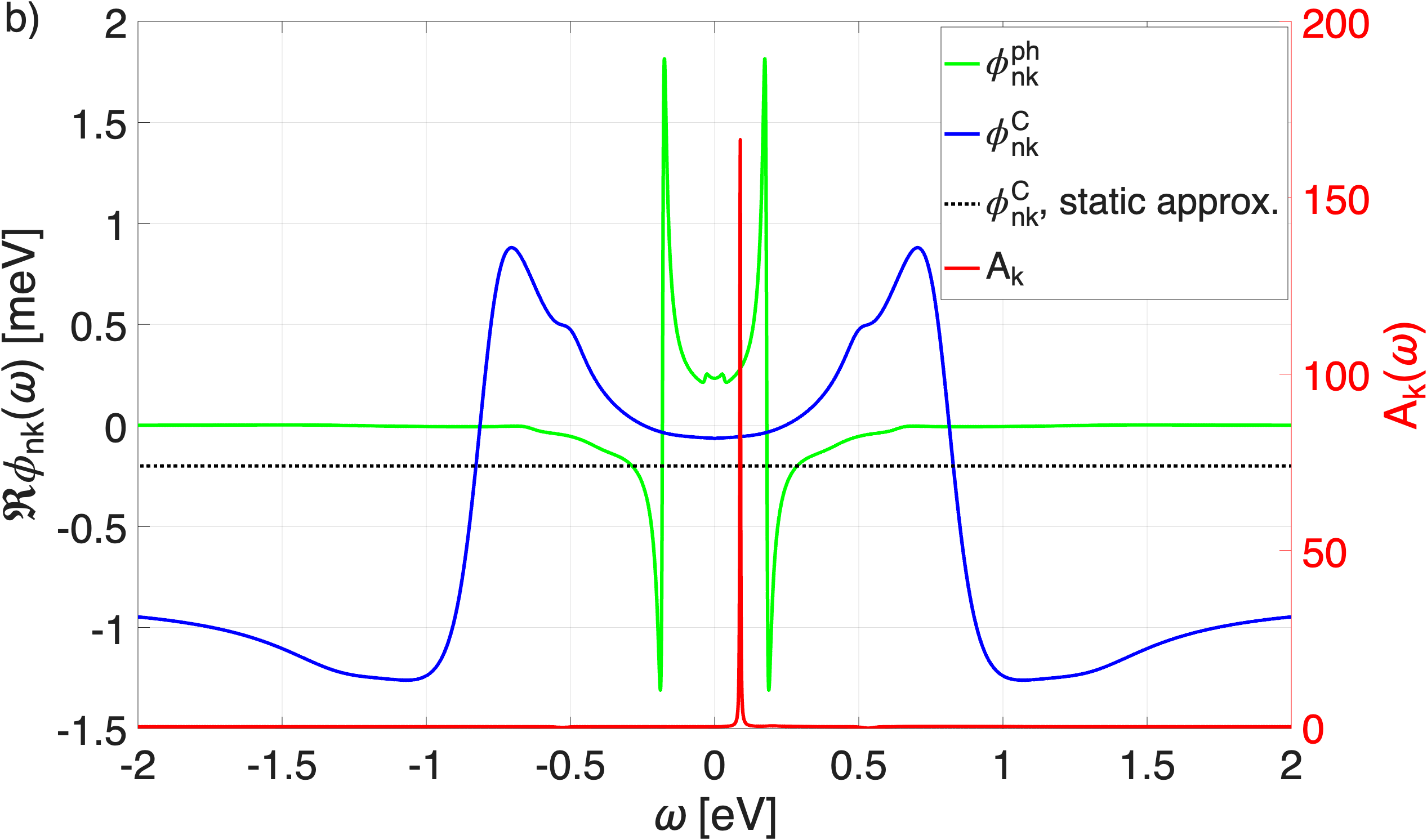}
\caption{a) The superconducting gap \red{at the Fermi level} of doped monolayer graphene calculated at $T=3$~K within s-GW, as a function of the energy cutoff along the Matsubara imaginary-frequency axis.
b) Anomalous (pairing) self-energy and spectral function (solid red line) for an electron with energy $E^{\rm QP}_{n\mathbf{k}} \approx 0.1$~eV above the Fermi level in doped monolayer graphene calculated within the s-GW at $T=3$~K. Solid green and blue lines represent the phonon and Coulomb contributions to the anomalous self-energy.
The dotted black horizontal line is the static‐screening value $\phi^{C \mathrm{, stat}}$, obtained using \eref{BCS_static} with all right‐hand‐side quantities taken from the converged s-GW solution.
%The prominent features around $\pm0.6$~eV in the Coulomb contribution $\phi^{C}$ correspond to the energy of the dominant coupling plasmon.
}
\label{scGW_graphene}
\end{figure}

Variations of the s-GW method have been widely employed to study superconductivity trends in multilayer graphene systems \cite{Lewandowski2021,Long2024,Veld2023}. Figure~\ref{scGW_graphene} presents our s-GW \textit{ab initio} results for doped monolayer graphene at $T=3$~K. Strikingly, 
the emergence of a nonzero superconducting gap $\Delta$ points to a critical temperature $T_c > 3$~K (see \fref{scGW_graphene}a), in direct contradiction  to experiments, which to date have found no superconducting transition in monolayer graphene down to mK temperatures. 

Figure~\ref{scGW_graphene}b shows the Coulomb‐induced anomalous self‐energy $\phi^C(\omega)$ on the real‐energy axis for a representative Bloch state $n \mathbf{k}$. The solid blue curve is the s-GW result, while the dotted black horizontal line is the static‐screening value $\phi^{C, \rm stat}$, defined as the left‐hand side of ~\eref{BCS_static} evaluated with all right‐hand‐side quantities taken from the converged s-GW solution.
At low energies, s-GW yields a negative 
$\phi^C(\omega)$, signifying net Coulomb repulsion, although its magnitude is reduced compared to the static‐screening value, so this regime remains pairing‐unfavorable. \red{
However, s-GW does promote pairing: as the energy increases beyond the phonon-mediated pairing window (indicated by the range over which the solid green curve, \(\phi^{\mathrm{ph}}(\omega)\), is significant), \(\phi^C(\omega)\) develops pronounced positive peaks around \(\pm 0.6\)~eV, associated with the dominant coupling plasmon. This high-energy attraction is precisely what drives the spurious overestimation of superconducting pairing and leads to the incorrect prediction of superconductivity in doped graphene. In particular, while the Coulomb contribution is repulsive at low energies, it develops pronounced positive peaks at higher energies, around the coupling-plasmon energy. Through the coupled Eliashberg equations, these high-energy attractive features are not confined to a single state \(n\mathbf{k}\), but can influence other coupled states and thereby promote the superconducting solution found within s-GW. Consistently, when the Coulomb channel is removed, we cannot find a superconducting solution.}

A previous study \cite{Davydov2020} that included the plasmon-mediated channel in the Eliashberg framework at the s-GW level found a significant overestimate of $T_c$ in bulk metals compared to the s-GW$_{static}$ approach.  We confirm the same limitation here for bulk Nb, as detailed below.
 
\begin{figure}
\includegraphics[trim=-0 0 -0 0,clip,width=\columnwidth]{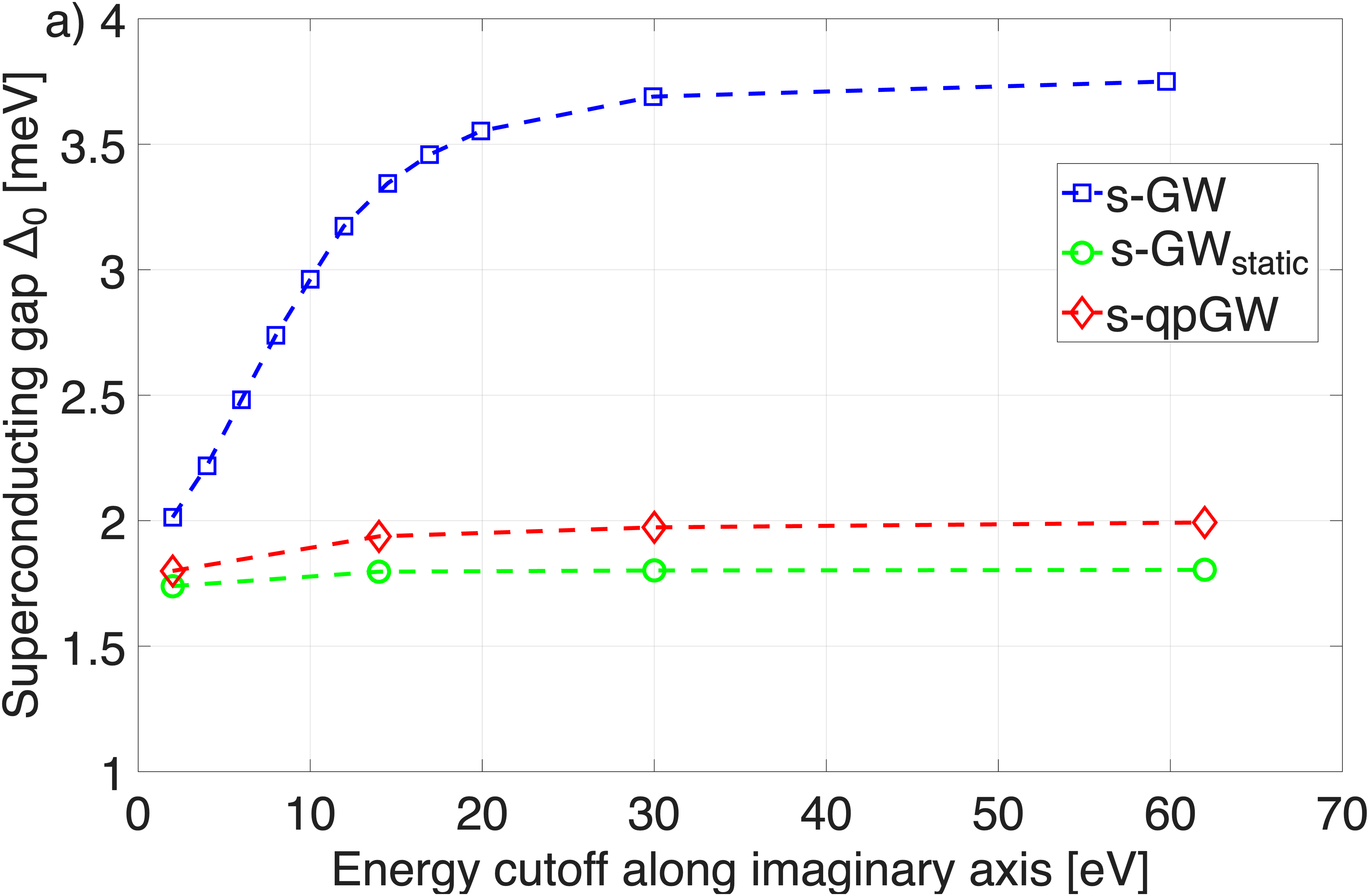}
\includegraphics[trim=-0 0 -0 0,clip,width=\columnwidth]{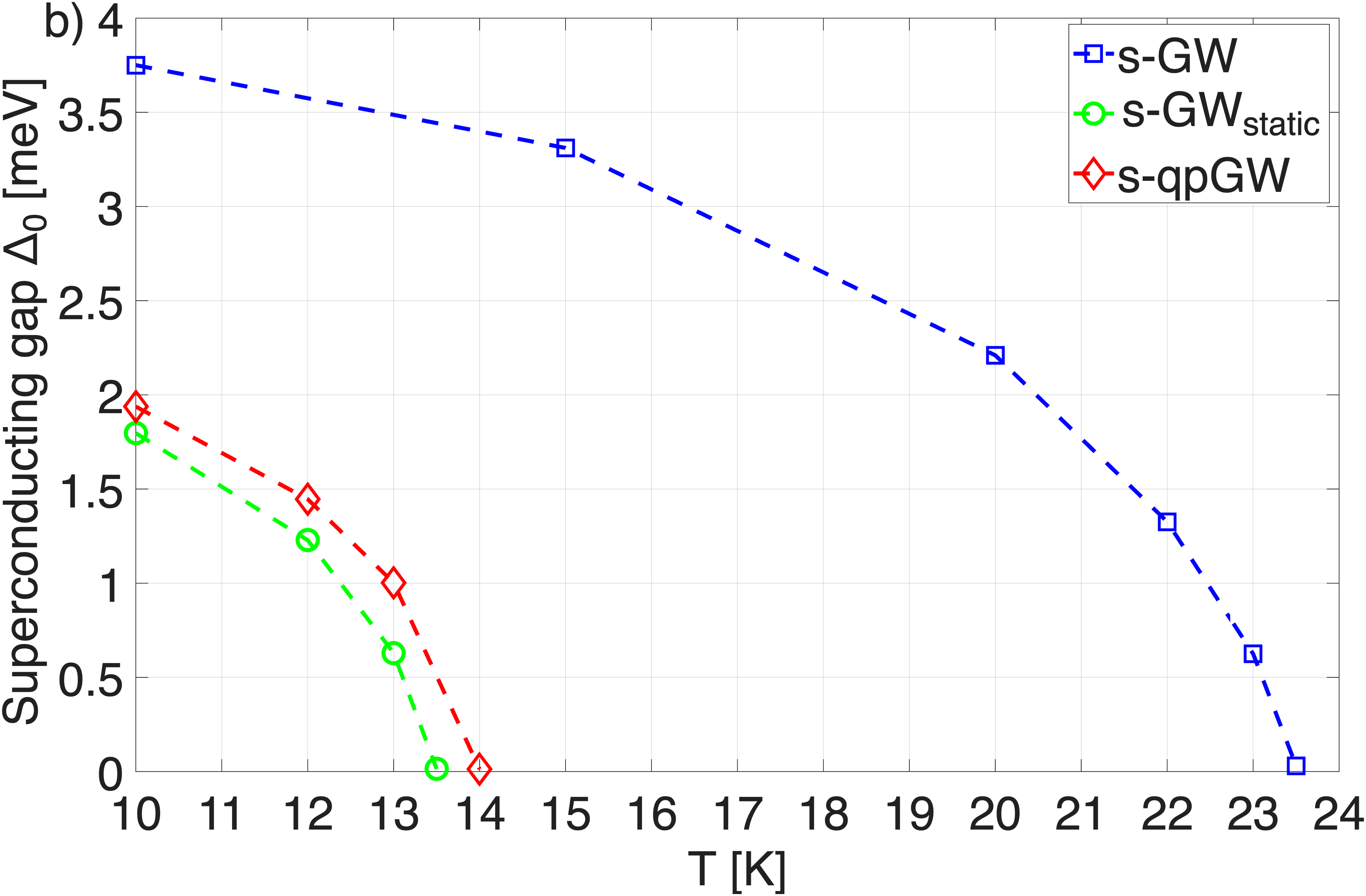}
 \caption{Superconducting properties of bulk Nb calculated
within s-GW (black squares), static Eliashberg (green circles), and s-qpGW (red diamonds). a) The superconducting gap \red{at the Fermi level} calculated at $T=10$~K, as function of the energy cutoff along the Matsubara imaginary-frequency axis. b) The converged superconducting gap as function of temperature.}
\label{Nb_Tc}
\end{figure}

We calculate the superconducting properties of bulk Nb, omitting spin‐fluctuation contributions, at each of the Coulomb‐channel levels of theory defined in  \sref{Formalism}. In particular, \fref{Nb_Tc}a shows the superconducting gap of Nb, calculated at $T=10$ K, within s-GW and the static Eliashberg approach, as a function of the energy cutoff 
%$\omega_{cut}^C$ 
along the Matsubara imaginary-frequency axis. Because s-GW induces spurious pairing from states far from the Fermi level, it leads to very slow convergence with respect to the energy cutoff, which must significantly exceed the bulk plasma frequency. To overcome this computational bottleneck, we employ a plasmon-pole approximation in conjunction with the Lindhard  dielectric model \cite{Mahan2000} for \(W^C(\omega)\). Further implementation details are provided in \appref{AppendixC}.

Figure~\ref {Nb_Tc}b shows the calculated superconducting gap (converged w.r.t. $\omega_c$), as function of temperature. The corresponding $T_c$ values for the static and s-GW cases are $13.5$~K and $23.5$~K, respectively. Our results within s-GW and s-GW$_{static}$  are in excellent agreement (within $< 0.5$~K) with existing literature \cite{Davydov2020,Mori2024}. When accounting for the missing  spin-fluctuation contributions, s-GW$_{static}$  agrees very well with experiment. 

One may ask why a more sophisticated theory level such as s-GW yields worse results than a more basic treatment of the Coulomb interactions, {\it i.e.} the static approximation?  This spurious enhancement traces back to self-consistency and the omission of vertex corrections in the electronic channel.  In fact, Migdal’s theorem tells us that vertex corrections can be safely neglected only when the characteristic bosonic (“glue”) energy $\omega_{bos}$ is small compared to the fermionic energy scale $E_F$, so that  $\frac{\omega_{bos}}{E_F} \ll 1\,$.
This condition is satisfied in conventional phonon-mediated superconductors where $\frac{\omega_{ph}}{E_F}\sim 10^{-2}$, and the standard Eliashberg (or BCS) treatment is controlled.  
By contrast, the plasmonic channel involves bosonic excitations at energies $\omega_{pl}$ that in bulk metals are several eV or more—comparable to the relevant electronic bandwidth in 3D systems.  In this regime $\frac{\omega_{pl}}{E_F}\sim 1$, the Migdal criterion breaks down, and omitting vertex diagrams overestimates the attractive part of the Coulomb interaction (as we observe in Nb and as previously noted in Ref.~\cite{Davydov2020}). In fact, for simple 3D jellium‐like metals the dynamically screened Coulomb channel predicts~\cite{Dassarma2025} critical temperatures on the order of $10^2–10^3$ K, manifestly at odds with experiment. As Das Sarma et al.~\cite{Dassarma2025} have shown, an uncritical Migdal-Eliashberg treatment of plasmon-mediated pairing in 3D metals is unjustified when the plasmon energy \(\omega_{pl}\) becomes comparable to the Fermi energy \(E_F\), since vertex corrections can no longer be neglected.
  
In strictly 2D systems the situation is more subtle. On the one hand, 2D plasmons are acoustic \red{in the sense that their energy vanishes in the long-wavelength limit (in graphene, \(\omega_{pl}(q)\propto\sqrt{q}\))}: \(\omega_{pl}(q\to0)\ll E_F\), and Migdal’s criterion is satisfied in this limit. On the other hand, for momentum transfers near the Fermi wavevector:
${\omega_{pl}(q\sim k_F)}\sim {E_F}$,
and the Migdal approximation breaks down.  
Indeed, our s-GW calculations for doped graphene (Fig.~\ref{scGW_graphene}a)
spuriously predict a finite superconducting gap (\(T_c>3\)\,K),
demonstrating that Migdal’s approximation breaks down in this 2D system.
Whether Migdal’s theorem can remain valid in other 2D materials
(e.g.\ twisted bilayer graphene) remains an open question.

\subsection{s-qpGW applied to doped graphene and Nb}

One natural solution for the missing‐vertex problem in self-consistent GW‐based theories comes from the electronic structure community’s use of the QP approximation. In particular, the QP self‐consistent GW method~\cite{Faleev2004,vanSchilfgaarde2006}, which combines the self‐consistent GW with the QP approximation, is a state‐of‐the‐art approach for normal‐state band‐structure calculations, correcting the pathologies of full self-consistent GW in describing the electronic structure.
To test whether this same QP‐projection remedy suppresses the spurious plasmon-mediated pairing generated in the superconducting phase by s-GW, we apply the s-qpGW scheme introduced in Sec.~\ref{subsec:s-qpGW} and Appendix~\ref{AppendixB}.

The recipe for implementing the s-qpGW method is straightforward and can be understood in the case of doped graphene by inspecting \fref{scGW_graphene}b.
Briefly, for an electronic state $n\mathbf{k}$, s-qpGW replaces the Coulomb superconducting order parameter with its real-part value evaluated at the superconducting QP energy: $\phi_{n\mathbf{k}}^{C}\left(\omega\right)\rightarrow \Re\phi_{n\mathbf{k}}^{C}(E_{n\mathbf{k}}^{\rm QP})$.  
\red{The relevant energy window is not a sharp adjustable cutoff, but depends on the relative importance of the bosonic modes that couple to the quasiparticles participating in pairing.}
Notably, for electrons within $\sim 0.2$ eV of the Fermi level (the phonon energy scale) and below the coupling‐plasmon energy ($\sim 0.6$ eV; \fref{scGW_graphene}b), the Coulomb anomalous self-energy remains negative, i.e. pairing-unfavorable, albeit with a smaller magnitude than in the static‐screening approximation (dotted black line). By contrast, the s-GW curve (solid blue line) turns positive outside the phonon window, driving the spurious pairing that leads to a nonzero superconducting gap. 
Within s-qpGW, the superconducting gap collapses to zero under self-consistent convergence down to 0.5 K.  Consequently, both s-qpGW and s-GW$_{static}$ predict no superconductivity in doped graphene, in agreement with the experiments.

\begin{figure}
\includegraphics[trim=-0 0 -0 0,clip,width=\columnwidth]{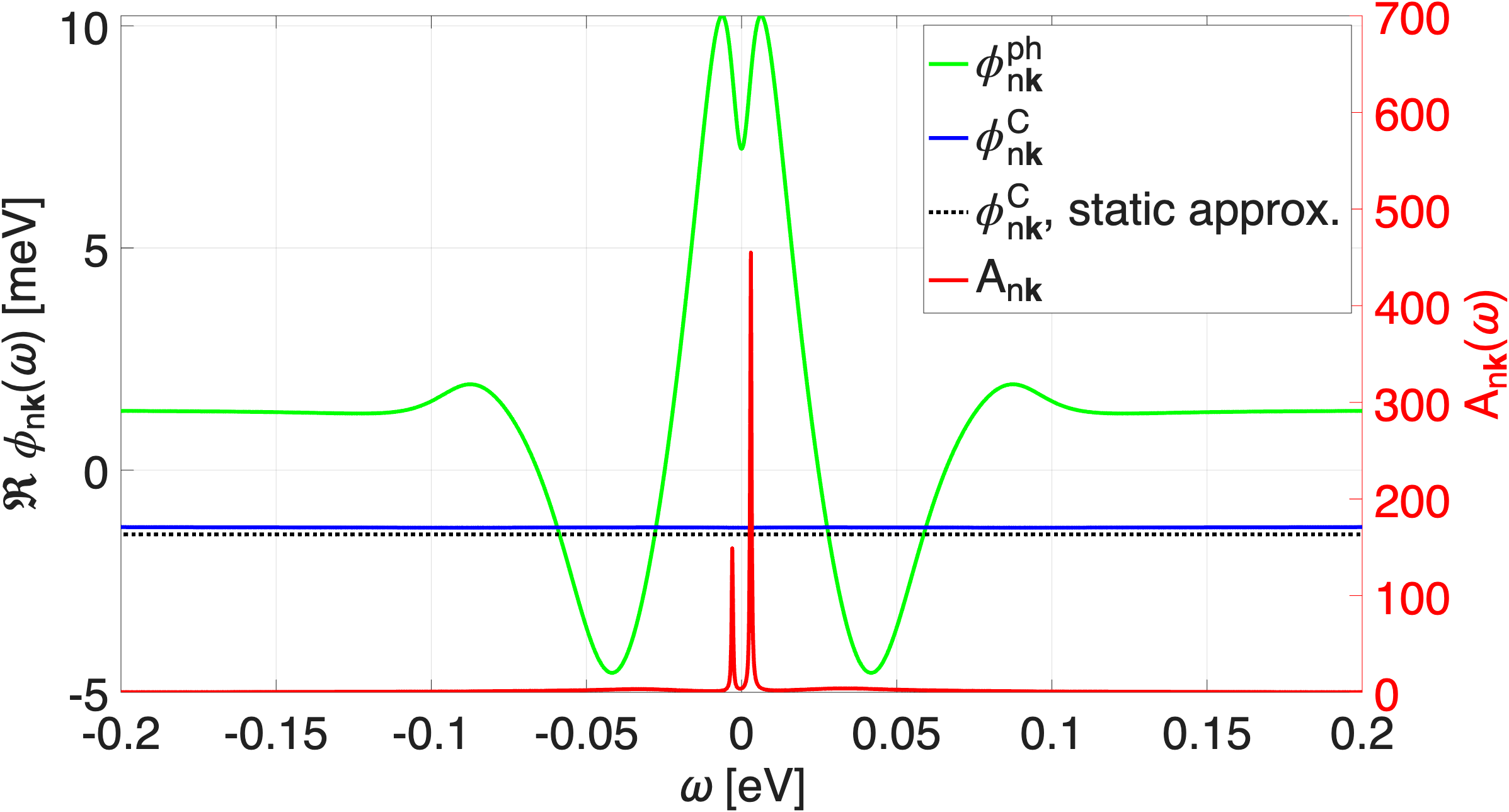}
\caption{Anomalous (pairing) self-energy and spectral function (solid red line)
for an electron with energy $E^{\rm QP}_{n\mathbf{k}} \approx 3$~meV above the Fermi level in bulk Nb, calculated within the s-qpGW at $T=10$~K.
Solid green and blue lines represent the phonon and Coulomb contributions to the anomalous self-energy.
The dotted black horizontal line is the static‐screening value $\phi^{C \mathrm{, stat}}$, obtained using \eref{BCS_static} with all right‐hand‐side quantities taken from the converged s-qpGW solution.
} 
\label{phi_A_Nb}
\end{figure}

For bulk Nb (see \fref{phi_A_Nb}), the plasmon energies are much higher than the phonon energy scale.
As a result, the Coulomb anomalous self‐energy is essentially flat over the phonon window:
$\Re\phi^{C,r}_{n\mathbf{k}}(\omega) \approx \phi^{C}_{n\mathbf{k}}(0)$.
In this case one can further approximate
$\tilde{\phi}^{C}_{n\mathbf{k}} \approx \phi^{C}_{n\mathbf{k}}(0)$.

Applying s-qpGW to bulk Nb yields excellent agreement with the prevailing state-of-the-art {\it ab initio} (static) Eliashberg theory \cite{Mori2024,Davydov2020}, which ignores dynamical screening effects.  This is illustrated in Fig.~\ref{Nb_Tc}b, where s-qpGW predicts $T_c=14$~K, within $0.5$~K of the static result. The slight offset reflects the fact that only in the limit $\omega_{pl}\rightarrow\infty$ 
 does the zero‐frequency Coulomb self‐energy 
$\phi^{C}_{n\mathbf{k}}(0)$ coincide with its static‐screening counterpart $\phi^{C,{stat}}_{n\mathbf{k}}$. In bulk Nb, where the plasmon energies far exceeds the phonon energies, this limit is almost realized.

Our calculations for bulk Nb and doped graphene show that s-qpGW removes the spurious plasmon-mediated pairing enhancement seen in s-GW, suggesting that s-qpGW can serve as a truly {\it ab initio} framework for dynamical screening in both 2D and 3D superconductors without compromising agreement with experiment. 

\subsection{s-qpGW applied to a model 2D system: graphene with an artificially enhanced density of states at $E_F$}

\begin{figure}
\includegraphics[trim=-0 0 -0 0,clip,width=\columnwidth]{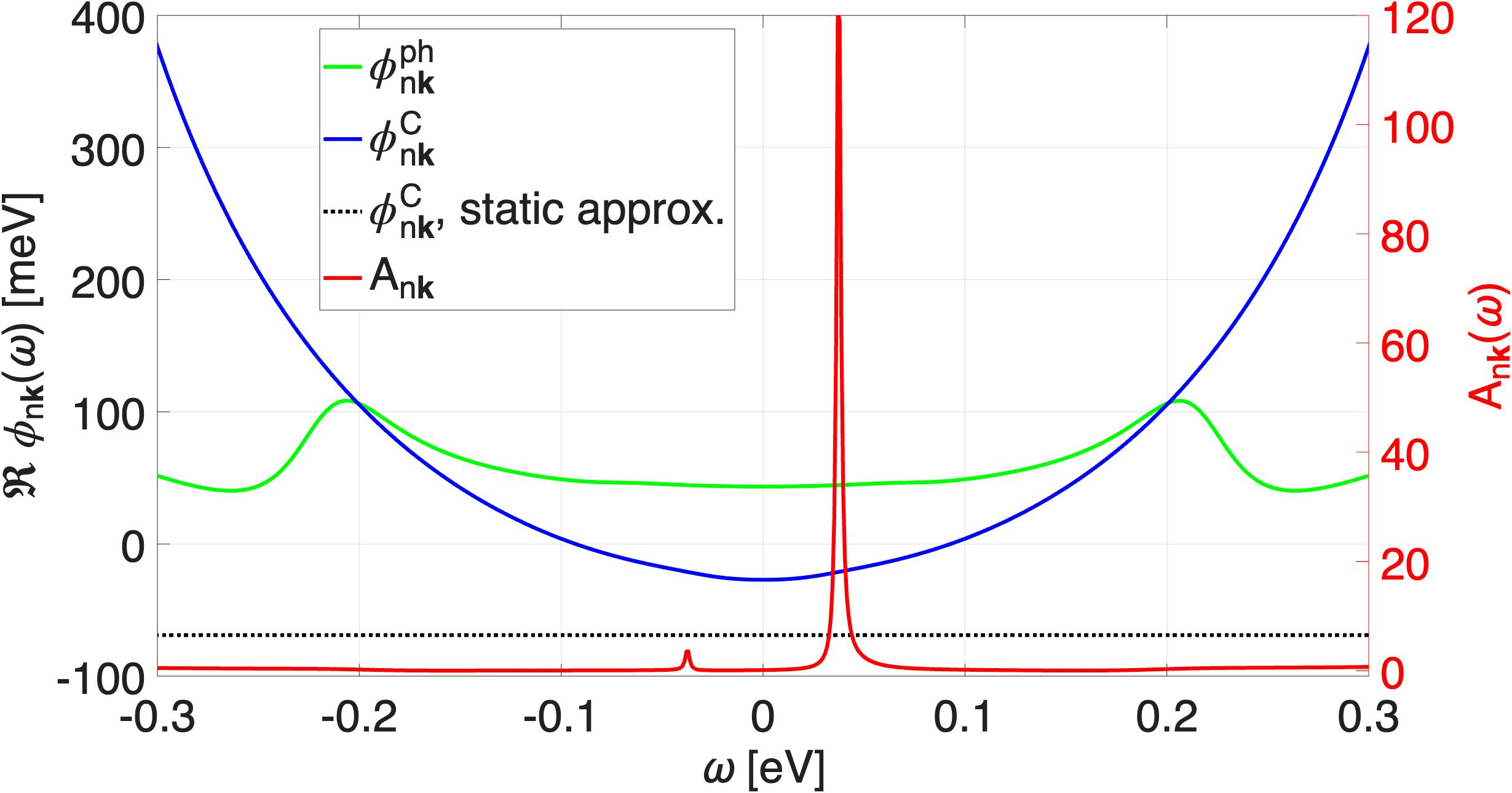}
\caption{Anomalous (pairing) self-energy and spectral function (solid red line)
for an electron with energy $E^{\rm QP}_{n\mathbf{k}} \approx 0.04$~eV above the Fermi level in doped graphene with an artificially enhanced ($\times20$) DOS(E$_F$), calculated within the s-qpGW at $T=10$~K.
Solid green and blue lines represent the phonon and Coulomb contributions to the anomalous self-energy.
The dotted black horizontal line is the static‐screening value $\phi^{C \mathrm{, stat}}$, obtained using \eref{BCS_static} with all right‐hand‐side quantities taken from the converged s-qpGW solution.}
\label{s-qpGW_gr}
\end{figure}

\begin{figure}
\includegraphics[trim=-0 0 -0 0,clip,width=\columnwidth]{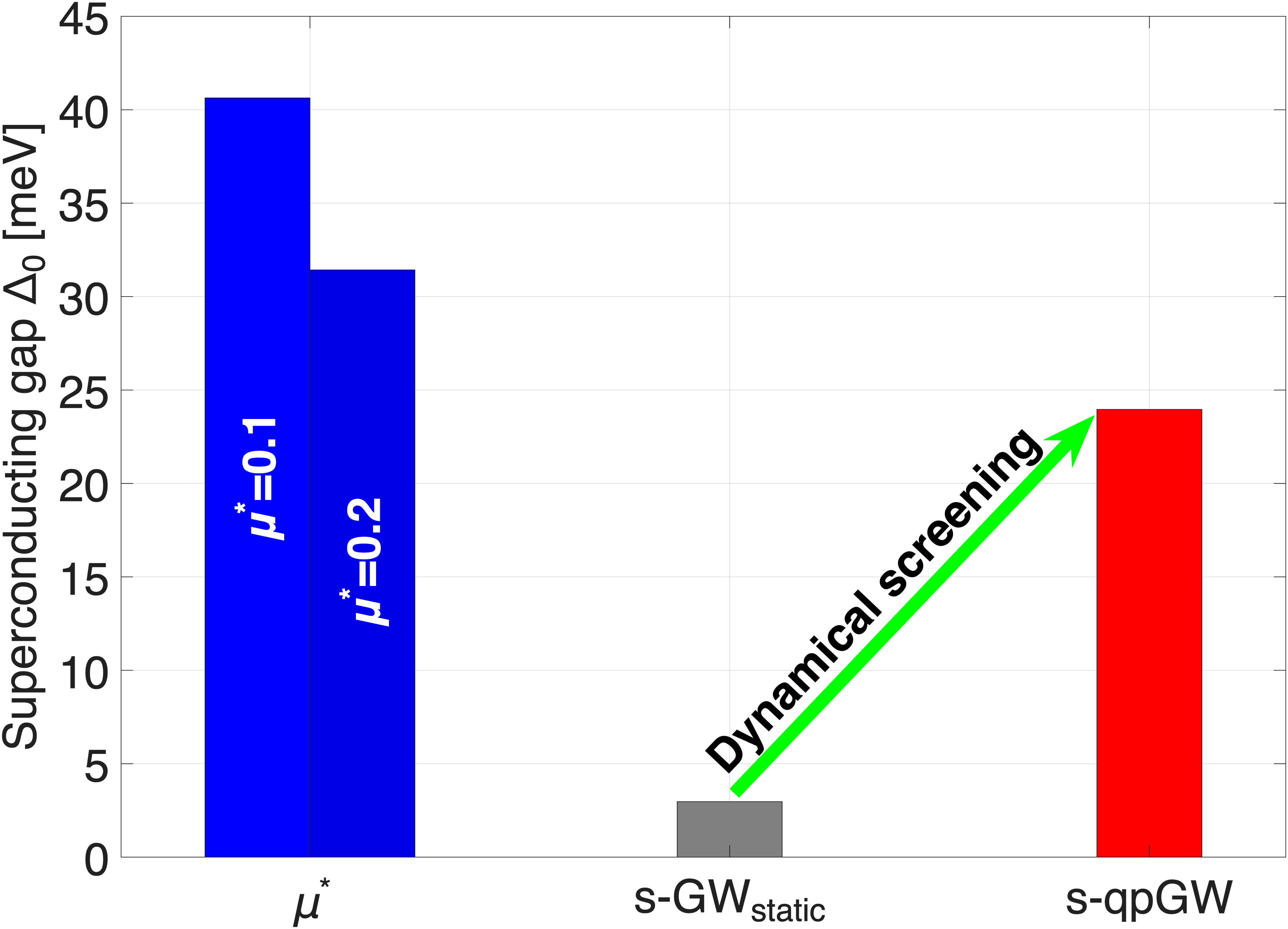}
\caption{Calculated superconducting gap \red{at the Fermi level} of doped graphene with an artificially enhanced ($\times20$) electronic density of states near the Fermi level, at $T=10$~K. The phonon contribution to $\Delta$ is fully included in all calculations. Results at the s-GW level are not shown as they grossly overestimate $\Delta$.}
\label{gr_DOSx20_Tc}
\end{figure}

\red{
To illustrate more clearly the effect of dynamical screening in a 2D setting, we consider a simple model system: doped graphene with an artificially enhanced density of states (DOS) at the Fermi level. Specifically, we multiply by $20$ the sums over Bloch states entering the Coulomb terms in Eqs.~\eqref{scGW_phiC} and \eqref{BCS_static} 
and the phonon terms in Eqs.~\eqref{eq17} and \eqref{eq15}.
This enhancement is motivated by prominent DOS features in other graphene systems, such as rhombohedral trilayer graphene, where superconductivity is found in proximity to van Hove singularities near the Fermi level \cite{Rubio24}.

Figure~\ref{s-qpGW_gr} shows the real part of the anomalous self-energy for the graphene model with a $\times 20$ enhanced DOS($E_F$), calculated within s-qpGW at $T=10\,$K for a quasiparticle $0.04$~eV above the Fermi level. In this case, plasmon energies can be comparable to phonon energies, and the Coulomb contribution is therefore no longer flat across the phonon window, unlike in bulk Nb. Comparing $\Re\phi^{C}_{n\mathbf{k}}(\omega)$ with the static-screening value $\phi^{C,\mathrm{stat}}$, we find that the latter is substantially more negative over the relevant low-energy range, which explains why s-GW$_{\mathrm{static}}$ underestimates the superconducting gap relative to s-qpGW (see also \fref{gr_DOSx20_Tc}). At the same time, the Coulomb term alone does not drive pairing in s-qpGW.
}

Figure~\ref{gr_DOSx20_Tc} shows the superconducting gap of our model system at $T= 10$ K. We compare three treatments of the Coulomb contribution to $\Delta$: i) The semi-empirical $\mu^*$ method; ii) Purely electrostatic screening within s-GW$_{static}$; and iii) Dynamical screening within s-qpGW. 

Comparing cases i) and ii), we note that the $\mu^*$ method, when employed with the conventional 3D parameter range 
$\mu^*_{3\mathrm{D}}\in[0.1,0.2]$ derived from bulk superconductors, deviates markedly from the standard static Eliashberg predictions in the 2D regime. While the magnitude of this deviation depends on the chosen DOS($E_F$) enhancement factor, these results are consistent with the lack of a formal justification for applying a semi-empirical $\mu^*$ calibrated on three‐dimensional materials to two‐dimensional systems \cite{Simonato2023}. Combined with the tendency of standard static Eliashberg theory to underestimate $T_c$ in few-layer graphene systems compared to experiments at the same doping level~\cite{SarmasTLG22},  these findings underscore the need for a theoretical framework capable of recalibrating $\mu^*$ specifically for 2D superconductors.

The pronounced discrepancy between cases ii) and iii) indicates that dynamical screening within s-qpGW can significantly influence predictions of superconducting properties in 2D systems. Indeed, in low-dimensional systems acoustic plasmons are ubiquitous and energetically compete with phonons, so neglecting their dynamical contribution may not be justified. 
\red{We believe that the reduction of the effective Coulomb repulsion can be captured qualitatively by a simple plasmon-pole form of the screened interaction. However, a fully transparent interpretation in terms of \(W^C(\omega)\) alone is difficult, because the effect becomes apparent only after self-consistency: upon convergence, the static and dynamical self-energies differ not only through the frequency dependence of the screened interaction, but also through the anomalous propagator entering Eq.~\eqref{Sigma_12}.}
We further find that choosing \(\mu^*=0.31\) brings the semi-empirical \(\mu^*\) method into quantitative agreement with s-qpGW. 
\red{
This larger value is qualitatively consistent with previous first-principles work on the layered, quasi-2D superconductor NbSe\(_2\) \cite{Sanna2022NbSe2}, where \(\mu^*=0.28\), rather than the usual assumed value \(\mu^*=0.11\), was needed to reproduce experiment. It is also broadly consistent with an earlier \(\mu^*\)-based study of trilayer graphene \cite{Rubio24}, where using \(\mu^*=0.2\) led to a substantial overestimate of \(T_c\). A possible interpretation is that the larger effective \(\mu^*\) reflects reduced screening efficiency in low-dimensional or quasi-2D systems relative to bulk 3D metals. We stress, however, that the analogy with our \(\times 20\) DOS graphene model should be interpreted cautiously, since it holds only to the extent that this artificial model captures the physics of a real low-dimensional material with enhanced DOS\((E_F)\).
}
Consequently, the s-qpGW method may also offer a theoretical framework for calibrating the \(\mu^*_{2D}\) range in 2D superconductors.

\section{Conclusion}
\label{Conclusion}

We have introduced the superconducting quasiparticle GW (s-qpGW) framework, a parameter-free extension of Eliashberg theory that incorporates dynamical Coulomb screening in systems of arbitrary dimensionality, similarly to its normal-phase counterpart. Unlike fully self-consistent GW, s-qpGW suppresses spurious plasmon-mediated pairing, reproducing the critical temperature $T_c$ predicted by standard (static) Eliashberg theory for bulk Nb and correctly predicting the absence of superconductivity in doped monolayer graphene. 

Moreover, s-qpGW captures dynamical Coulomb screening by acoustic plasmons, thereby suppressing the pairing-breaking electrostatic repulsion, an effect demonstrated in a model 2D system where s-qpGW predicts superconducting gaps that substantially exceed those from the static Eliashberg approach. Further validation of s-qpGW through applications to prototypical two-dimensional materials, such as few-layer graphene, remains an important future direction. 

We anticipate that the s-qpGW framework will provide new insights into the superconducting mechanisms of graphene-based and other 2D materials (including transition-metal dichalcogenides \cite{TMDs1,TMDs2}), particularly regarding the interplay between electron-phonon and electron-plasmon interactions. These insights should help close the knowledge gap in 2D superconductivity theory and enhance our ability to control and optimize the performance of superconducting electronic devices built from these materials.

\red{\section{Methods}
\label{Methods}}

\red{The first-principles calculations are performed with the Quantum ESPRESSO package~\cite{Giannozzi2017} with optimized norm-conserving Vanderbilt pseudopotentials (ONCVPSPs)~\cite{Hamann2013, Schlipf2015}. We use the local density approximation (LDA)~\cite{Perdew1992} for graphene and the Perdew-Burke-Ernzerhof (PBE) parametrization of the generalized gradient approximation~\cite{Perdew1996} for Nb. The plane wave cutoff is set to 100~Ry in both cases.  A graphene layer in isolation is described using a supercell geometry with periodic replicas separated by 20~\text{\AA} in the $z$-direction. The optimized lattice parameter for $n$-doped graphene with 0.05 electrons per unit cell is $a=2.455$~\text{\AA}, and for Nb is $a = 3.309$~\text{\AA}. The electron charge density is computed on a $72\times72\times1$ \textbf{k} mesh for graphene, and a $16\times16\times16$ \textbf{k} mesh for Nb, with a Methfessel-Paxton smearing~\cite{Methfessel1989} of 0.10~eV. The dynamical matrices and the linear variation of the self-consistent potential are calculated within density-functional perturbation theory~\cite{Baroni2001} on the irreducible set of a regular $12\times12\times1$ \textbf{q} mesh for graphene, and $8\times8\times8$ \textbf{q} mesh for Nb.

The maximally localized Wannier functions used for the Wannier-Fourier interpolation in EPW are constructed using the WANNIER90 code~\cite{Marzari1997,Pizzi2020,Marrazzo2024}. For graphene, five Wannier functions are used to describe the
electronic states near the Fermi level: \(sp^2\)-like orbitals associated with the in-plane \(\sigma\) manifold and \(p_z\)-like orbitals centered on the C atoms. For Nb, we use nine Wannier functions with \(s\), \(p\), and \(d\) character.

We employ the EPW code~\cite{Ponce2016,Lee2023,Margine2013}  to solve the anisotropic Migdal-Eliashberg equations. The electronic eigenenergies, phonon frequencies, and electron-phonon matrix elements are evaluated on \(120\times120\times1\)
$\mathbf{k-}$ and $\mathbf{q-}$ meshes for graphene, and \(36\times36\times36\) $\mathbf{k-}$ and $\mathbf{q-}$ meshes for Nb. The Fermi windows are set to $\pm 0.5$ eV and $\pm 0.3$ eV around $E_F$ for graphene and Nb, respectively. }
\\

\begin{acknowledgments}
We thank Wei Pan and Feliciano Giustino for useful discussions.
The authors acknowledge support from the Laboratory Directed Research and
Development program at Sandia National Laboratories. 
Sandia National Laboratories is a multi-mission laboratory 
managed and operated by National
Technology and Engineering Solutions of Sandia, LLC, a
wholly owned subsidiary of Honeywell International, Inc.,
for the U.S. DOE’s National Nuclear Security Administration 
under contract DE-NA-0003525. The views expressed in the article do not necessarily represent the
views of the U.S. DOE or the United States Government.
C. R. and E. R. M. acknowledge support from the National
Science Foundation (NSF) under awards no. OAC-2103991 and OAC-2513830.  
\end{acknowledgments}

\renewcommand{\theequation}{A-\arabic{equation}} 
\setcounter{equation}{0}
\setcounter{subsection}{0}

\appendix
\section{Phonon contributions $\phi^{ph}$ and $Z^{ph}$}
%\section*{Appendix A: The phonon contribution to $\phi$ and $Z$.}
\label{AppendixA}

%Starting from \eref{Wph-D} and transforming to the real-space phonon normal modes (the eigenfunctions, labeled by wavevector \(q\) and branch index \(\nu\), of the lattice dynamical matrix):
Starting from \eref{Wph-D}, the phonon contribution to the screened Coulomb 
interaction in real space is written as a double convolution of the Coulomb interaction with the phonon propagator
\begin{align}
W^{ph}(\mathbf r,\mathbf r';\omega)
&=\int d\mathbf r_{1}\,\int d\mathbf r_{2}\;
  W^{C}(\mathbf r,\mathbf r_{1})\, \nonumber \\
& \times  
  D(\mathbf r_{1},\mathbf r_{2};\omega)\,
  W^{C}(\mathbf r_{2},\mathbf r').
%  \tag{1}\\
\label{eq:Wph_realspace}
\end{align}
The phonon propagator is expanded in the eigenfunctions $ \eta_{q\nu}(\mathbf r)$ of the lattice dynamical matrix $D_{\mathbf{q}\nu}(\omega)$, labeled by wave vector $\mathbf{q}$ and branch index $\nu$
\begin{align}
D(\mathbf r_{1},\mathbf r_{2};\omega)
&=\sum_{\mathbf{q}\nu}
  \eta_{\mathbf{q}\nu}(\mathbf r_{1})\,
  D_{\mathbf{q}\nu}(\omega)\,
  \eta^{*}_{\mathbf{q}\nu}(\mathbf r_{2}).
\label{eq:D_modeexp}
\end{align}
Inserting \eqref{eq:D_modeexp} into \eqref{eq:Wph_realspace} yields
\begin{align}
W^{ph}(\mathbf r,\mathbf r';\omega)
&=\sum_{\mathbf{q}\nu}D_{\mathbf{q}\nu}(\omega) \nonumber \\
& \times
  \!\int\!d\mathbf r_{1}\,W^{C}(\mathbf r,\mathbf r_{1})\,\eta_{\mathbf{q}\nu}(\mathbf r_{1}) \nonumber\\
& \times   
  \!\int\!d\mathbf r_{2}\,\eta^{*}_{\mathbf{q}\nu}(\mathbf r_{2})\,W^{C}(\mathbf r_{2},\mathbf r') \nonumber\\
&=\sum_{\mathbf{q}\nu}
  g_{\mathbf{q}\nu}(\mathbf r)\,
  D_{\mathbf{q}\nu}(\omega)\,
  g^{*}_{\mathbf{q}\nu}(\mathbf r')\,,
 \label{eq:Wph_gDg}
\end{align}

\noindent
where we have defined the electron-phonon vertex in the normal‐mode basis as
\begin{equation}
g_{\mathbf{q}\nu}(\mathbf r)
\;=\;
\int d\mathbf r_{1}\;W^{C}(\mathbf r,\mathbf r_{1})\;\eta_{\mathbf{q}\nu}(\mathbf r_{1})\,.
\end{equation}

The phonon propagator \(D_{\mathbf{q}\nu}(\omega)\) admits the standard representation
\beq
D_{\mathbf{q}\nu}(\omega)
=\frac{2\,\omega_{\mathbf{q}\nu}}{\omega^{2}-\omega_{\mathbf{q}\nu}^{2}+i0^{+}}.
\eeq

Projecting $W^{ph}(\mathbf r,\mathbf r';\omega)$ onto Bloch states
$\psi_{n\mathbf{k}}(\mathbf r)$, we obtain
\beq
W^{ph}_{n\mathbf{k},n'\mathbf{k}'}(\omega)
= \sum_{\nu}
  \bigl\lvert g_{n\mathbf{k},n'\mathbf{k}'}^{\nu}\bigr\rvert^{2}\,
  D_{\mathbf{k}-\mathbf{k}',\nu}(\omega),
\label{W-ph-kspace}
\eeq
where $g_{n\mathbf{k},n'\mathbf{k}'}^{\nu}$ is the electron-phonon matrix
element for scattering between the electronic states $(n\mathbf{k})$ and
$(n'\mathbf{k}')$ mediated by a phonon with wave vector
$\mathbf q = \mathbf k - \mathbf k'$ and branch index $\nu$,
\beq
g_{n\mathbf{k},n'\mathbf{k}'}^{\nu}
= \int d\mathbf r\,
  \psi^{*}_{n'\mathbf{k}'}(\mathbf r)\,
  g_{\mathbf q\nu}(\mathbf r)\,
  \psi_{n\mathbf{k}}(\mathbf r).
\eeq

The anisotropic electron-phonon coupling is defined as
\begin{align}
\lambda_{n\mathbf{k},n'\mathbf{k}'}
      (i\omega_j - i\omega_{j'})
&= \int_{0}^{\infty} d\omega\;
   \frac{2\,\omega}
        {(\omega_{j}-\omega_{j'})^{2}+\omega^{2}} \nonumber\\
&\quad\times
   \alpha^{2}F_{n\mathbf{k},n'\mathbf{k}'}(\omega),
\label{eq:lambda}
\end{align}
with the usual Eliashberg spectral function
\begin{equation}
\alpha^{2}F_{n\mathbf{k},n'\mathbf{k}'}(\omega)
= N_F \sum_{\nu}
  \bigl\lvert g_{n\mathbf{k},n'\mathbf{k}'}^{\nu} \bigr\rvert^{2}
  \delta\!\bigl(\omega-\omega_{\mathbf{k}-\mathbf{k}',\nu}\bigr).
\label{eq:alpha2F}
\end{equation}

Comparing Eqs.~\eqref{W-ph-kspace} and \eqref{eq:lambda}, and using the Matsubara
form of the phonon propagator
\beq
D_{\mathbf q\nu}(i\omega_j - i\omega_{j'}) =
- \frac{2\omega_{\mathbf q\nu}}{\bigl[(\omega_j-\omega_{j'})^{2}
+\omega_{\mathbf q\nu}^{2}\bigr]}, 
\eeq
we find 
\beq
\lambda_{n\mathbf{k},n'\mathbf{k}'}
      (i\omega_j - i\omega_{j'})
= - N_F\,
  W^{ph}_{n\mathbf{k},n'\mathbf{k}'}
      (i\omega_j - i\omega_{j'}) .
\eeq

Using Eqs.~\eqref{phi_def} and \eqref{Z_def}, the phonon contributions to
$\phi_{n\mathbf{k}}(i\omega_{j})$ and $Z_{n\mathbf{k}}(i\omega_{j})$ can be
written as 
\begin{equation}
\begin{split}
\phi^{ph}_{n\mathbf{k}}(i\omega_{j})
&= {T}
   \sum_{n'\mathbf{k}' j'}
   \frac{\phi_{n'\mathbf{k}'}(i\omega_{j'})}
        {\det\hat{G}^{-1}_{n'\mathbf{k}'}(i\omega_{j'})} \\[2pt]
&\quad\times      
   \,
   W^{ph}_{n\mathbf{k},n'\mathbf{k}'}
      (i\omega_j - i\omega_{j'}) ,
\end{split}
\label{eq17}
\end{equation}
\begin{equation}
\begin{split}
Z^{ph}_{n\mathbf{k}}(i\omega_{j})
&= \frac{T}{\omega_{j}}
   \sum_{n'\mathbf{k}' j'}
   \frac{\omega_{j'}\,Z_{n'\mathbf{k}'}(i\omega_{j'})}
        {\det\hat{G}^{-1}_{n'\mathbf{k}'}(i\omega_{j'})} \\[2pt]
&\quad\times      
   \,
   W^{ph}_{n\mathbf{k},n'\mathbf{k}'}
      (i\omega_j - i\omega_{j'}) .
\end{split}
\label{eq15}
\end{equation}

\renewcommand{\theequation}{B-\arabic{equation}} 
\setcounter{equation}{0}
\setcounter{subsection}{0}
\section{Derivation of the s-qpGW method}
\label{AppendixB}

In the normal (non‐superconducting) state, the quasiparticle self-consistent GW method (QSGW) \cite{Faleev2004,vanSchilfgaarde2006}—which employs a QP approximation to compensate for the missing vertex corrections in standard self-consistent GW—has proven reliable for predicting band structures and related electronic properties of weakly to moderately correlated materials. In this Appendix we extend the QP approximation to the superconducting phase.
We emphasize that the derivation does not depend on QSGW per se; one may equally start from an eigenvalue-only self-consistent GW reference, either evGW or its partially self-consistent variant evGW\(_0\) \cite{Kresse07,Shih2010}, and recover the same working equations.  In fact, evGW\(_0\) in the normal state is directly analogous to the superconducting QP approximation implemented here: both keep the screened interaction \(W\) fixed at its initial RPA value and freeze the QP wavefunctions at their underlying Kohn–Sham forms.

\subsection*{Quasiparticle Self-Consistent GW for the Normal State}
We begin by summarizing QSGW’s treatment of the normal-state self-energy. In QSGW the frequency dependence of the electronic self-energy,  
$\Sigma(\omega)\equiv\Sigma^{C}_{11}(\omega)$, is removed via the QP approximation. In the Bloch basis $\{\ket{n\mathbf{k}}\}$, we define
\beq
\begin{aligned}[b]
\bra{n\mathbf{k}}\tilde{\Sigma}\ket{m\mathbf{k}}
&= \frac{1}{2}
   \Big[
   \bra{n\mathbf{k}}\Re \Sigma(\epsilon_{n\mathbf{k}})\ket{m\mathbf{k}} \\
&\quad ~ +
   \bra{n\mathbf{k}}\Re \Sigma(\epsilon_{m\mathbf{k}})\ket{m\mathbf{k}}
   \Big],
\end{aligned}
\label{sigma_QSGW}
\eeq
where $\mathbf{k}$ denotes the crystal momentum, $n$ and $m$ are band indices,
$\epsilon_{n\mathbf{k}}$ and $\epsilon_{m\mathbf{k}}$ are the QP energies, and
$\Re$ in front of an operator denotes its Hermitian part,
\beq
\Re \Sigma(\omega) \equiv \frac{\Sigma^{r}(\omega)+\Sigma^{a}(\omega)}{2}.
\eeq
Here $\Sigma^{r,a}(\omega)$ are the retarded and advanced self-energies, obtained by analytic continuation from the Matsubara imaginary-frequency axis to the real-frequency axis via $i\omega_{j} \rightarrow \omega\pm i\eta$.

QSGW can be cast as a symmetrized convolution with the QP spectral function
$A_{n\mathbf{k}}(\omega)$:
\beq
%\begin{split}
\begin{aligned}[b]
\bra{n\mathbf{k}}\tilde{\Sigma}\ket{m\mathbf{k}}
  &= \frac{1}{2}
     \int d\omega\,
     \bra{n\mathbf{k}}\Re\Sigma(\omega)\ket{m\mathbf{k}}
     A_{m\mathbf{k}}(\omega)  \\ %[4pt]
  \quad & ~+ \frac{1}{2}
     \int d\omega\,
     A_{n\mathbf{k}}(\omega)
     \bra{n\mathbf{k}}\Re\Sigma(\omega)\ket{m\mathbf{k}},
\end{aligned}
\label{QSGW_norm}
\eeq
where
\beq
A_{n\mathbf{k}}(\omega)
= \frac{G^{a}_{n\mathbf{k}}(\omega)-G^{r}_{n\mathbf{k}}(\omega)}{2\pi i},
\eeq
and the quasiparticle Green’s functions are
\beq
G^{r/a}_{n\mathbf{k}}(\omega)
= \frac{1}{\omega-\epsilon_{n\mathbf{k}}\pm i0^{+}}.
\eeq

\subsection*{Extending Quasiparticle Self-Consistent GW to Superconductors}

We now extend \eref{QSGW_norm} to the superconducting state in
Nambu space, using as usual \cite{Margine2013} the diagonal approximation (neglecting band mixing), and
treating everything as $2 \times 2$ matrices. For each $(n,\mathbf{k})$ the
static Coulomb self-energy is defined as a symmetrized convolution with the
Nambu spectral function,
\beq
\begin{aligned}[b]
\hat{\tilde{\Sigma}}^{C}_{n\mathbf{k}}
&= \frac{1}{2} 
   \int d\omega\,
        \Re\hat{\Sigma}^{C}_{n\mathbf{k}}(\omega)\,
        \hat{A}_{n\mathbf{k}}(\omega) \\ %[5pt]
\quad &+ \frac{1}{2} 
   \int d\omega\,
        \hat{A}_{n\mathbf{k}}(\omega)\,
        \Re\hat{\Sigma}^{C}_{n\mathbf{k}}(\omega),
\end{aligned}
\label{s-qpGW_supercond}
\eeq
where
\beq
\hat{A}_{n\mathbf{k}}(\omega)
= \frac{\hat{G}^{a}_{n\mathbf{k}}(\omega)
      - \hat{G}^{r}_{n\mathbf{k}}(\omega)}{2\pi i},
\eeq
and $\hat{G}^{r/a}_{n\mathbf{k}}(\omega)$ are the full Nambu Green's
functions (containing both phonon and electron self-energy contributions).
The spectral function obeys the standard Nambu relations, 
\beq
\begin{split}
&[\hat A_{n\mathbf{k}}(\omega)]_{12} = [\hat A_{n\mathbf{k}}(\omega)]_{21}, \\
&[\hat A_{n\mathbf{k}}(\omega)]_{22} = [\hat A_{n\mathbf{k}}(-\omega)]_{11}.
\end{split}
\eeq

The lack of frequency dependence in $\hat{\tilde{\Sigma}}^{C}_{n\mathbf{k}}$
ensures that the corresponding mass renormalization function is zero and thus
we write
\beq
\hat{\tilde{\Sigma}}^{C}_{n\mathbf{k}} =
\begin{pmatrix}
\tilde{\chi}^{C}_{n\mathbf{k}} & \tilde{\phi}^{C}_{n\mathbf{k}} \\
\tilde{\phi}^{C}_{n\mathbf{k}} & -\tilde{\chi}^{C}_{n\mathbf{k}}
\end{pmatrix},
\eeq
where $\tilde{\chi}^{C}_{n\mathbf{k}}$ and $\tilde{\phi}^{C}_{n\mathbf{k}}$
are real, frequency-independent Coulomb contributions to the normal and
anomalous self-energies.

The frequency-dependent Coulomb self-energy entering
Eq.~\eqref{s-qpGW_supercond} has the general Nambu form
\beq
\begin{aligned}[b]
\Re\hat{\Sigma}^{C}_{n\mathbf{k}}(\omega) =
\begin{pmatrix}
\Re[\Sigma^{C}_{n\mathbf{k}}(\omega)]_{11} & \Re\phi^{C}_{n\mathbf{k}}(\omega) \\
\Re\phi^{C}_{n\mathbf{k}}(\omega) & \Re[\Sigma^{C}_{n\mathbf{k}}(\omega)]_{22}
\end{pmatrix},
\end{aligned}
\eeq
where, according to Eqs.~\eqref{Sigma_11_Z_chi}, \eqref{Sigma_22_Z_chi} and \eqref{scGW_ZC},
\beq
\begin{aligned}[b]
\Re[\Sigma^{C}_{n\mathbf{k}}(\omega)]_{11/22}
= - \omega\,\Re Z^{C}_{n\mathbf{k}}(\omega)
  \pm \Re\chi^{C}_{n\mathbf{k}}(\omega).
\end{aligned}
\eeq

Equation~\eqref{s-qpGW_supercond} allows us to identify the two components of
$\hat{\tilde{\Sigma}}^{C}_{n\mathbf{k}}$:
\beq
\begin{aligned}[b]
\tilde{\chi}^{C}_{n\mathbf{k}}
&= \int d\omega\,
     \Re[\Sigma^{C}_{n\mathbf{k}}(\omega)]_{11}\,
     [\hat A_{n\mathbf{k}}(\omega)]_{11} \\
&\quad
  + \int d\omega\,
     \Re\phi^{C}_{n\mathbf{k}}(\omega)\,
     [\hat A_{n\mathbf{k}}(\omega)]_{12},
\end{aligned}
\label{chi_tilde}
\eeq
\beq
\begin{aligned}[b]
\tilde{\phi}^{C}_{n\mathbf{k}}
&= \frac{1}{2} \int d\omega\,
     \Re\phi^{C}_{n\mathbf{k}}(\omega)\, \\
& \quad \times     
     \Big(
       [\hat A_{n\mathbf{k}}(\omega)]_{11}
      +[\hat A_{n\mathbf{k}}(\omega)]_{22}
     \Big) \\
&\quad
  + \frac{1}{2} \int d\omega\,
     \Big(  
     \Re[\Sigma^{C}_{n\mathbf{k}}(\omega)]_{11}
     +\Re[\Sigma^{C}_{n\mathbf{k}}(\omega)]_{22} 
     \Big)   \\
& \quad \times         
     [\hat A_{n\mathbf{k}}(\omega)]_{12}.
\end{aligned}
\label{phi_tilde}
\eeq

The spectral function $[\hat A_{n\mathbf{k}}(\omega)]_{11}$ has a two-peak
structure at the positive and negative superconducting QP energies
$\pm E^{\rm QP}_{n\mathbf{k}}$. The peak weights are the product of the
quasiparticle renormalization constant
$Z^{\rm QP}_{n\mathbf{k}}
 =[1-d\Re[\Sigma_{n\mathbf{k}}(\omega)]_{11}/d\omega]^{-1}_{|\omega=E^{\rm QP}_{n\mathbf{k}}}$
and the BCS-like coherence factors $u^{2}_{n\mathbf{k}}$ and $v^{2}_{n\mathbf{k}}$, which satisfy
$u_{n\mathbf{k}}^2 \;+\; v_{n\mathbf{k}}^2 \;=\; 1.$
Within s-qpGW, the deviation of \(Z^{\mathrm{QP}}_{n\mathbf{k}}\) from unity arises solely from electron--phonon coupling.
The anomalous spectral function $[\hat A_{n\mathbf{k}}(\omega)]_{12}$ also has
peaks at $\pm E^{\rm QP}_{n\mathbf{k}}$ and is odd in frequency.
Noting that
$[\hat A_{n\mathbf{k}}(\omega)]_{11} + [\hat A_{n\mathbf{k}}(-\omega)]_{11}$ is an even
function of $\omega$ (with peaks at 
$\pm E^{\rm QP}_{n\mathbf{k}}$), Eqs.~\eqref{chi_tilde} and \eqref{phi_tilde} simplify if we make use of the even parity
under $\omega \rightarrow -\omega$ of
$\Re\phi^{C}_{n\mathbf{k}}(\omega)$, $\Re Z^{C}_{n\mathbf{k}}(\omega)$, and
$\Re\chi^{C}_{n\mathbf{k}}(\omega)$ [a consequence of the corresponding scalar
Matsubara quantities $\phi^{C}_{n\mathbf{k}}(i\omega)$,
$Z^{C}_{n\mathbf{k}}(i\omega)$, and $\chi^{C}_{n\mathbf{k}}(i\omega)$ being
real-valued and even under $i\omega \rightarrow -i\omega$], and of the
odd parity of $[\hat A_{n\mathbf{k}}(\omega)]_{12}$. We obtain
\beq
\begin{aligned}[b]
\tilde{\chi}^{C}_{n\mathbf{k}}
&= - \int_{0}^{\infty} d\omega\,
     \omega\,\Re Z^{C}_{n\mathbf{k}}(\omega)\, \\
& \quad \times     
     \Big(
       [\hat A_{n\mathbf{k}}(\omega)]_{11}
      -[\hat A_{n\mathbf{k}}(-\omega)]_{11}
     \Big)
\\
&\quad
  + \int_{0}^{\infty} d\omega\,
     \Re\chi^{C}_{n\mathbf{k}}(\omega)\, \\
& \quad \times     
     \Big(
       [\hat A_{n\mathbf{k}}(\omega)]_{11}
      +[\hat A_{n\mathbf{k}}(-\omega)]_{11}
     \Big),
\end{aligned}
\label{chi_tilde_bis}
\eeq
\beq
\begin{aligned}[b]
\tilde{\phi}^{C}_{n\mathbf{k}}
&= \int_{0}^{\infty} d\omega\,
     \Re\phi^{C}_{n\mathbf{k}}(\omega)\, \\
& \quad \times     
     \Big(
       [\hat A_{n\mathbf{k}}(\omega)]_{11}
      +[\hat A_{n\mathbf{k}}(-\omega)]_{11}
     \Big)
\\
&\quad
  - 2 \int_{0}^{\infty} d\omega\,
      \omega\,\Re Z^{C}_{n\mathbf{k}}(\omega)\,
      [\hat A_{n\mathbf{k}}(\omega)]_{12}.
\end{aligned}
\label{phi_tilde_bis}
\eeq

\red{
It is worth noting that in the non-superconducting limit, \eref{chi_tilde_bis} gives \(\tilde{\phi}^{C}_{n\mathbf{k}}=0\), as expected. If one further approximates the normal spectral function by a single quasiparticle pole at \(\epsilon_{n\mathbf{k}}\), consistent with the QP approximation, \eref{phi_tilde_bis}  reduces to
\beq
\begin{aligned}[b]
\tilde{\chi}^{C}_{n\mathbf{k}} &\approx
-\epsilon_{n\mathbf{k}}\,\Re Z^{C}_{n\mathbf{k}}(\epsilon_{n\mathbf{k}})
+\Re\chi^{C}_{n\mathbf{k}}(\epsilon_{n\mathbf{k}})\\
&=
\Re[\Sigma^{C}_{n\mathbf{k}}(\epsilon_{n\mathbf{k}})]_{11},
\end{aligned}
\label{non-SC_lim}
\eeq
which recovers \eref{sigma_QSGW} in the band-diagonal approximation.

Returning to the superconducting phase, we neglect as usual \cite{Margine2013} the normal component and set \(\tilde{\chi}^{C}_{n\mathbf{k}}=0\).
Two s-qpGW flavors can then be considered, depending on whether
\(\Re Z^{C}_{n\mathbf{k}}(\omega)\) is neglected or not.\\

i) The most straightforward implementation neglects \(\Re Z^{C}_{n\mathbf{k}}(\omega)\), in which case
\eref{phi_tilde_bis} reduces to
\beq
\begin{aligned}[b]
\tilde{\phi}^{C}_{n\mathbf{k}}
&= \int_{0}^{\infty} d\omega\,
     \Re\phi^{C}_{n\mathbf{k}}(\omega)\, \\
& \quad \times
     \Big(
       [\hat A_{n\mathbf{k}}(\omega)]_{11}
      +[\hat A_{n\mathbf{k}}(-\omega)]_{11}
     \Big),
\end{aligned}
\label{phi_tilde_this_work_a}
\eeq
which is equivalent to
\beq
\tilde{\phi}^{C}_{n\mathbf{k}}
= \int d\omega\,
\Re\phi^{C}_{n\mathbf{k}}(\omega)\,
[\hat A_{n\mathbf{k}}(\omega)]_{11}.
\label{phi_tilde_this_work_b}
\eeq

Finally, neglecting the incoherent part of the spectral weight (i.e.\ setting
\(Z^{\rm QP}_{n\mathbf{k}}\approx1\)), and consistent with the QP approximation, one approximates
\[
[\hat A_{n\mathbf{k}}(\omega)]_{11}+[\hat A_{n\mathbf{k}}(-\omega)]_{11}
\approx \delta(\omega-E^{\rm QP}_{n\mathbf{k}})
+\delta(\omega+E^{\rm QP}_{n\mathbf{k}}),
\]
which leads directly to \eref{phi_tilde_main},
\beq
\tilde{\phi}^{C}_{n\mathbf{k}}
\approx \Re\phi^{C}_{n\mathbf{k}}\!\left(E^{\mathrm{QP}}_{n\mathbf{k}}\right).
\label{phi_tilde_QP_result}
\eeq
This is the s-qpGW flavor used throughout this work.\\

ii) One may alternatively retain \(\Re Z^{C}_{n\mathbf{k}}(\omega)\) and directly use
\eref{phi_tilde_bis}. Within the QP approximation, the anomalous spectral function
\([\hat A_{n\mathbf{k}}(\omega)]_{12}\) is also represented by two poles at
\(\pm E^{\rm QP}_{n\mathbf{k}}\), with spectral weight \(u_{n\mathbf{k}}v_{n\mathbf{k}}\) for each peak, where
\(u_{n\mathbf{k}}\) and \(v_{n\mathbf{k}}\) are the usual coherence factors. Equation~\eref{phi_tilde_bis} then gives
\beq
\begin{aligned}[b]
\tilde{\phi}^{C}_{n\mathbf{k}}
&\approx \Re\phi^{C}_{n\mathbf{k}}\!\left(E^{\mathrm{QP}}_{n\mathbf{k}}\right)
\\
&\quad
-2E^{\rm QP}_{n\mathbf{k}}\,u_{n\mathbf{k}}v_{n\mathbf{k}}\,
\Re Z^{C}_{n\mathbf{k}}\!\left(E^{\mathrm{QP}}_{n\mathbf{k}}\right).
\end{aligned}
\label{phi_tilde_with_ZC}
\eeq
Using
\[
u_{n\mathbf{k}}v_{n\mathbf{k}}
=
\frac{\Delta_{n\mathbf{k}}}{2E^{\rm QP}_{n\mathbf{k}}},
\]
\[
\Delta_{n\mathbf{k}}
=
\frac{
\Re \phi^{\rm ph}_{n\mathbf{k}}(E^{\rm QP}_{n\mathbf{k}})
+
\Re \phi^{C}_{n\mathbf{k}}(E^{\rm QP}_{n\mathbf{k}})
}{
1+\Re Z^{\rm ph}_{n\mathbf{k}}(E^{\rm QP}_{n\mathbf{k}})
+\Re Z^{C}_{n\mathbf{k}}(E^{\rm QP}_{n\mathbf{k}})
},
\]
this may be rewritten as
\beq
\tilde{\phi}^{C}_{n\mathbf{k}}
\approx
\Re\phi^{C}_{n\mathbf{k}}\!\left(E^{\mathrm{QP}}_{n\mathbf{k}}\right)
-
\Delta_{n\mathbf{k}}\,
\Re Z^{C}_{n\mathbf{k}}\!\left(E^{\mathrm{QP}}_{n\mathbf{k}}\right).
\label{phi_tilde_with_ZC_gap}
\eeq

We have not explored the s-qpGW flavor ii) in the present work. We note that  in the limiting case where the electron-phonon couplings are absent, this expression reduces to
\[
\tilde{\phi}^{C}_{n\mathbf{k}}
\approx
\frac{\Re\phi^{C}_{n\mathbf{k}}(E^{\rm QP}_{n\mathbf{k}})}
{1+\Re Z^{C}_{n\mathbf{k}}(E^{\rm QP}_{n\mathbf{k}})},
\]
i.e.\ it differs from the flavor i) expression in \eref{phi_tilde_QP_result} by the multiplicative factor \(1/[1+\Re Z^{C}_{n\mathbf{k}}(E^{\rm QP}_{n\mathbf{k}})]\).
}
%%%

%\newpage
\renewcommand{\theequation}{C-\arabic{equation}} 
\setcounter{equation}{0}
\setcounter{subsection}{0}
\section{Implementation Details of s-GW for Nb}
\label{AppendixC}

We start from \eref{scGW_phiC} and, for simplicity, suppress all band and momentum indices, along with the associated sums.:
\begin{equation}
\begin{split}
\phi^{C}(i\omega_{j})
&= T
   \sum_{j'}
   \frac{\phi_{}(i\omega_{j'})}
        {\det\hat{G}^{-1}(i\omega_{j'})}   \,W^{C}(i\omega_j-i\omega_{j'}) ,
\end{split}
\end{equation}
We now split the fermionic Matsubara sum at a cutoff $\omega_{cut}^C$$>\omega_{cut}^{ph}$, such that the phonon contribution $\phi^{ph}(i\omega_{cut}^{C})\approx 0$, and the frequency dependence of Coulomb term $\phi^{C}$ is neglected and approximated as $ \phi^{C}(i\omega_{j})\approx \phi^{C}(i\omega_{cut}^C)\equiv\phi^{C}_{\rm cut} \quad \text{for} ~\omega_{j}>\omega_{cut}^C$. \eref{wcutC} then follows:
\beq
\begin{aligned}[b]
\phi^{C}&(i\omega_{j})
 = T
   \sum_{j'}^{\omega_{j'}<\omega_{cut}^C}
   \frac{\phi_{}(i\omega_{j'})-\phi^{C}_{\rm cut}}
        {\det\hat{G}^{-1}(i\omega_{j'})}   \,W^{C}(i\omega_j-i\omega_{j'}) \\
&+ T\sum_{j'}
   \frac{\phi^{C}_{\rm cut}}
        {\bigl(i\omega_{j'}\bigr)^{2} - \epsilon^2-(\phi^{C}_{\rm cut})^2}   
        \,W^{C}(i\omega_j-i\omega_{j'}).
\end{aligned}
\label{wcutC}
\eeq

$W^C$ is evaluated at momentum $q$ and bosonic Matsubara frequency $i\nu\equiv i\omega_j-i\omega_{j'}$ using the spectral representation
\beq
\begin{aligned}[b]
W^C(q,i\nu) =
    v(q) \Bigl[1 + \frac{2}{\pi}
       \int_{0}^{\infty} \!d\omega\,
       \frac{\omega\,\Im\,\epsilon^{-1}(q,\omega)}
            {\nu^{2} + \omega^{2}}\Bigr]\,, 
\end{aligned}
\eeq
where $v(q)$ is the bare Coulomb potential. Finally, we replace this by a simple plasmon-pole model together with the Lindhard form \cite{Lindhard_note} for the low-$T$ limit of the static dielectric function $\epsilon(q,0)$,
\begin{align}
W^C(q,i\nu) =
    v(q) \Bigl[1 - \frac{\Omega_{pl}^2}{\nu^{2} + \omega_{pl}^{2}(q)}\Bigr]
\,,
\label{PPM}
\end{align}
with $\omega_{pl}^{2}(q)=\Omega_{pl}^{2}/\bigl[1-\epsilon^{-1}(q,0)\bigr]$ and $\Omega_{pl}$ the plasma frequency of bulk Nb.
\red{
We verified that, for lower cutoff values \(\omega_{cut}^C\), the superconducting gap obtained with the plasmon-pole representation is in very good agreement with that obtained from the corresponding full-frequency Lindhard treatment.
}
% eps_min1_Mats_PPM(iZ) = 1.d0-E_plasma_eV**2/((omegaZ(iZ)/eV)**2+w2_PPM_eV)
% w2_PPM_eV=E_plasma_eV**2/(1.d0-1.d0/eps(1)) ! eq. 1.52 my thesis

\eref{wcutC}, together with the plasmon-pole model of \eref{PPM}, permits an analytic evaluation of the second term on its right-hand side.  This in turn accelerates convergence with respect to the cutoff~\(\omega_{cut}^C\), as illustrated by the blue line in \fref{Nb_Tc}(a).  Without this modification—i.e.\ setting $\phi^{C}_{\rm cut}=0$ in \eref{wcutC}—convergence with respect to \(\omega_{cut}^C\) becomes prohibitively slow, rendering the calculation impractical.

%%%%%%%%%%%%%%%%Biblipography%%%%%%%%%%%
\bibliography{refs}

\end{document}